\documentclass[aps,nofootinbib,reprint,groupedaddress]{revtex4-2}
\usepackage{amsmath,amssymb}
\usepackage{bbold}
\usepackage{float}
\usepackage{graphicx}% Include figure files
\usepackage{tikz}
\usepackage{dcolumn}% Align table columns on decimal point
\usepackage{bm}% bold math
\usepackage[T1]{fontenc}
\usepackage[colorlinks]{hyperref}

\usepackage{atbegshi}

\usepackage{color}
\definecolor{TQgreen1}{RGB}{89, 169, 123}
\definecolor{TQgreen2}{RGB}{37,	127,	75}

\begin{document}

\title{\color{TQgreen2} Practical Application-Specific Advantage through\\ Hybrid Quantum Computing}

\author{
{\color{black} Michael Perelshtein}}

\author{
{\color{black} Asel Sagingalieva}}

\author{
{\color{black} Karan Pinto}}

\author{
{\color{black} Vishal Shete}}

\author{
{\color{black} Alexey Pakhomchik}}

\author{
{\color{black} Artem Melnikov}}

\author{
{\color{black} Florian Neukart}}

\author{
{\color{black} Georg Gesek}}

\author{
{\color{black} Alexey Melnikov}}

\author{
{\color{black} Valerii Vinokur}}

\affiliation{Terra Quantum AG, 9400 Rorschach, Switzerland}
\affiliation{QMware AG, 9400 Rorschach, Switzerland}

\maketitle

    \tikz [remember picture, overlay] %
    \node [shift={(-1.7cm,0.1cm)}] at (current page.south east) %
    [anchor=south east] %
    {\includegraphics[width=15mm]{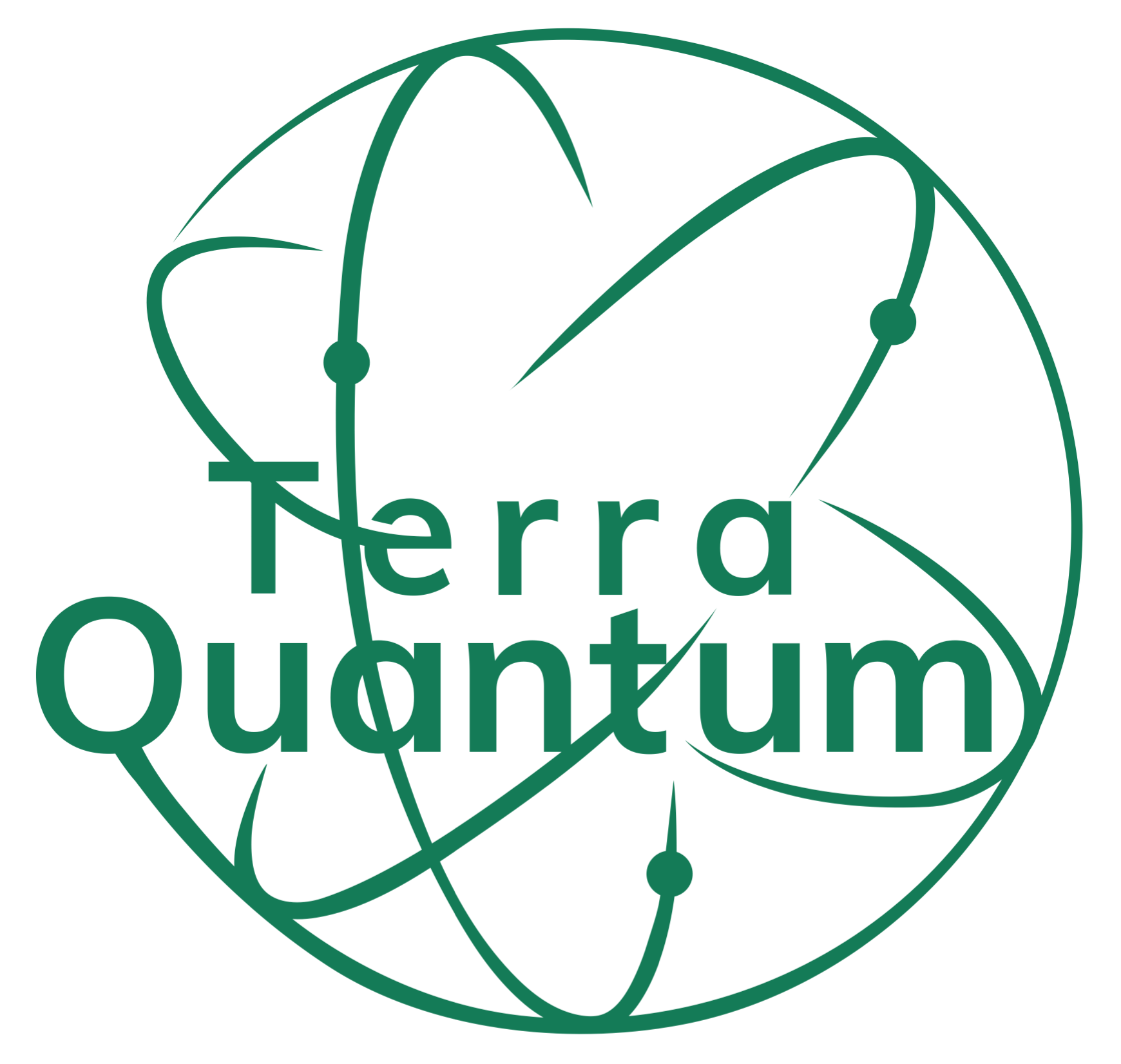}};

{\fontfamily{qtm}\selectfont
{\bf 
Quantum computing promises to tackle technological and industrial problems insurmountable for classical computers. However, today’s quantum computers still have limited demonstrable functionality, and it is expected that scaling up to millions of qubits is required for them to live up to this touted promise. The feasible route in achieving practical quantum advantage goals is to implement a hybrid operational mode that realizes the cohesion of quantum and classical computers. Here we present a hybrid quantum cloud based on a memory-centric and heterogeneous multiprocessing architecture, integrated into a high-performance computing data center grade environment. We demonstrate that utilizing the quantum cloud, our hybrid quantum algorithms including Quantum Encoding (QuEnc), Hybrid Quantum Neural Networks and Tensor Networks enable advantages in optimization, machine learning, and simulation fields. We show the advantage of hybrid algorithms compared to standard classical algorithms in both the computational speed and quality of the solution. The achieved advance in hybrid quantum hardware and software makes quantum computing useful in practice today.
}
}

{\color{TQgreen2}
\section{Introduction}}

Explosive development of quantum technologies imposes a challenge to correctly identify the approach most effective in exploiting the potential of quantum computing and successful in addressing the required industry-relevant problems in full-scale. The present state of the quantum art is attested as Noisy Intermediate-Scale Quantum (NISQ) technology, in which quantum computers comprising 50-100 qubits are on their way of surpassing the capabilities of today’s classical digital computers but in which quantum device decoherence, measurement imperfections, control errors and architectural limitations retard the further size growth of quantum circuits that can be reliably implemented~\cite{Preskill2018,bharti2022noisy}. The most promising way to achieve the desirable development of computing technology on the practical applications level, is to hybridize the powers of the available NISQ devices with state-of-the-art classical high-performance computing capabilities~\cite{bravyi2016trading,mcclean2016theory,li2017hybrid,zhu2019training}. This hybridization is needed not only for combining the powers of the different types of processors, quantum processing units (QPUs), central processing units (CPUs) and graphical processing units (GPUs), but also for a classical control, optimization, calibration, and error-correction of NISQ devices. Classical devices are important for verification of quantum devices~\cite{gheorghiu2019verification,eisert2020quantum}, and for providing logical quantum devices~\cite{semnanian2011virtualization,bechtold2021bringing}. The task is thus to integrate all components into a single platform, so that classical and quantum computing units exchange information locally and allow for efficient high-speed device-to-device and device-to-memory connections.

\begin{figure}[h!]
    \centering
    \includegraphics[width=1\linewidth]{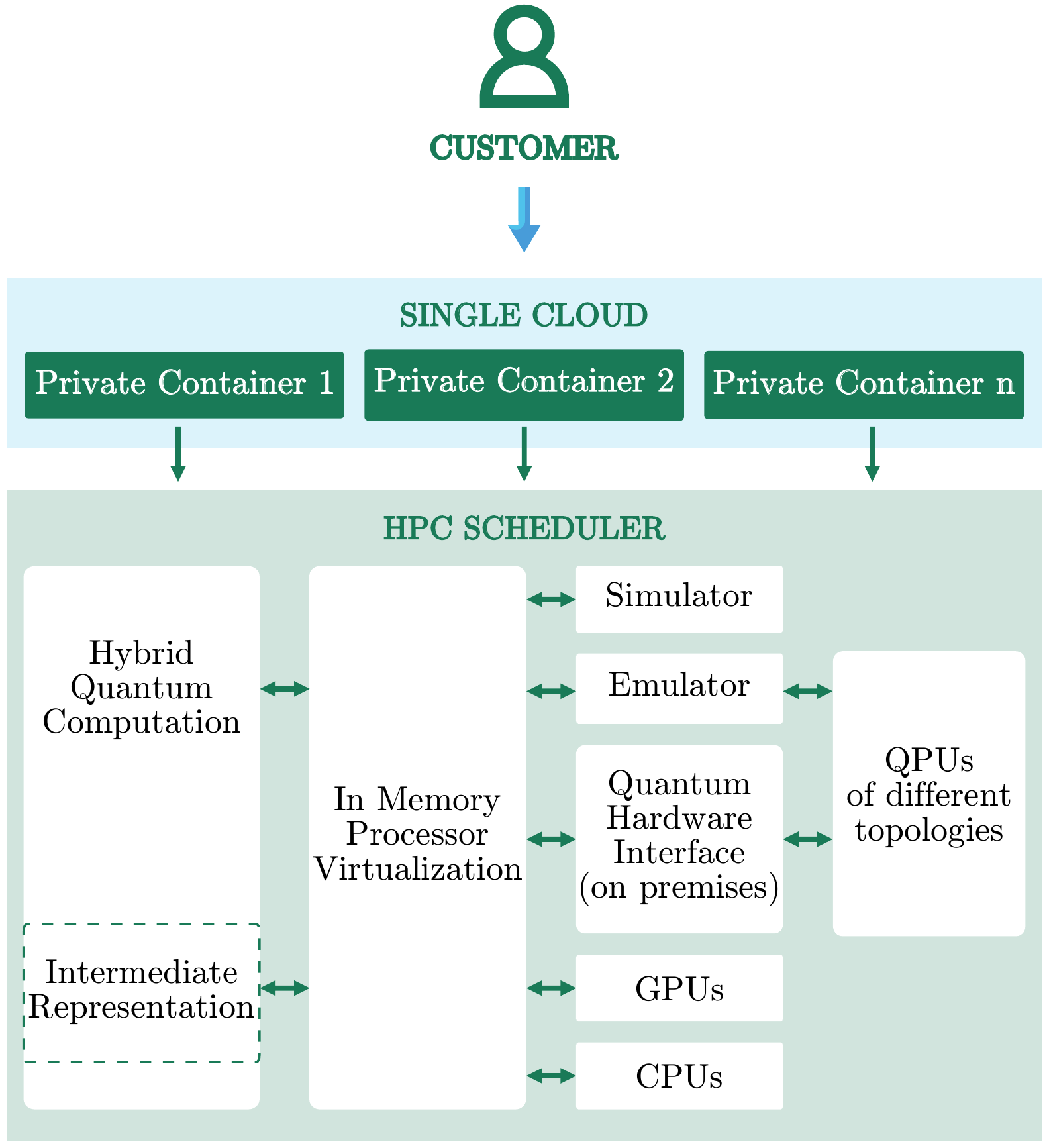}
    \caption{QMware hybrid quantum cloud: architecture diagram. A customer accesses hardware resources via one of the private containers.}
    \label{fig:HCloud1}
\end{figure}

Here we develop a Hybrid Quantum High-Performance Computing (HQC) cloud, with an in-memory computing model and complete software stack including hardware, operating system, middleware, and application programming interfaces (API), that fuses together the best of today's classical computing elements and the emerging quantum components into a hybrid model for productive use. Our hybrid quantum cloud, QMware, that we introduce here, provides a private cloud-based platform and containerized environment in a memory-centric compute architecture where research partners and industrial organizations can build and deploy their own hybrid quantum applications at scale~\cite{QMWare}. Our state-of-the-art GAIA-X-compliant data centers combine high-performance classical infrastructure, simulated QPUs~\cite{Guerreschi2020,mandra2021hybridq,ATOS,efthymiou2021qibo} and advanced machine learning tools for efficient production grade deployment of hybrid quantum-classical algorithms. There are three key innovations of the QMware cloud, which are shown schematically in Fig.~\ref{fig:HCloud1}. First, a memory-centric compute architecture, instead of the von Neumann architecture~\cite{von1993first,mariantoni2011implementing}, is used. It supports heterogeneous processing with QPUs, CPUs and GPUs as well as hybrid quantum compute, which includes all  processes and procedures involved in a hybrid quantum computation of a hybrid quantum application in a container. Second, a unified information theoretical model for both, classical and quantum information, represented as the memory patterns in our central main memory and integrated into our operating system. This model allows for efficient QPU simulation through implementing a hardware type agnostic intermediate representation of the quantum circuits (see Fig.~\ref{fig:HCloud1}, and e.g., Ref.~\cite{QIR}). The Intermediate Representation here reflects a generic model of the quantum circuits as part of the hybrid quantum algorithms. Our HPC and Quantum Simulation workloads are powered by Intel Xeon Platinum CPUs and NVIDIA A100 GPUs. Third, a unified in memory communication protocol, which makes sure that the right hardware accesses the relevant information pattern.

Here we focus on the hybrid algorithms, demonstrating how the combination of classical and simulated quantum resources realizes hybrid-exclusive advantages in optimization, machine learning, and simulation. In optimization, we demonstrate that our hybrid solution to the MaxCut problem showcases a performance advantage over the commercial high-performance CPLEX solver. In machine learning, we show the application-specific advantage of the hybrid quantum neural networks over their classical counterparts by studying two standard problems in the classification and regression domains. In simulation, we exhibit that the quantum-inspired tensor network differential equations' solution is more scalable than a classical conjugate gradient solution.

The paper is structured as follows. First, we introduce the hybrid quantum cloud and describe its capabilities to run hybrid quantum algorithms. Next, we demonstrate applications of hybrid quantum algorithms to optimization, machine learning, and simulation. In the summary, we discuss the impact of our developments.

\bigskip

{\color{TQgreen2}
\section{Hybrid Quantum Computing}
}

To build optimal full-scale hybrid solutions, we focus on implementing quantum algorithms in a high-performance classical environment, while evaluating the scaling on NISQ devices. While the speed-up of many quantum algorithms can be demonstrated analytically, e.g., Shor's algorithm~\cite{Shor}, most algorithm developing initiatives benefit considerably from numerical experiments implemented within sufficiently powerful environments. For instance, variational algorithms for quantum chemistry, optimization, and machine learning that are particularly adapted for NISQ devices~\cite{VQC2021} also involve a classical heuristic optimization routine that must be analysed numerically. The key ingredients of the emerging hybridized quantum computing directions are as follows:
\smallskip
\begin{enumerate}
    \item Combining optimally quantum and classical hardware resources for algorithm execution, within the given limitations of both units. \item Implementing quantum solutions with classical improvements to eliminate bottlenecks, and using classical algorithms for optimizing quantum subroutines.
    \item Synchronizing the processing and storage of the quantum and classical information across the stack
\end{enumerate}

\begin{figure}[h!]
    \centering
    \includegraphics[width=1\linewidth]{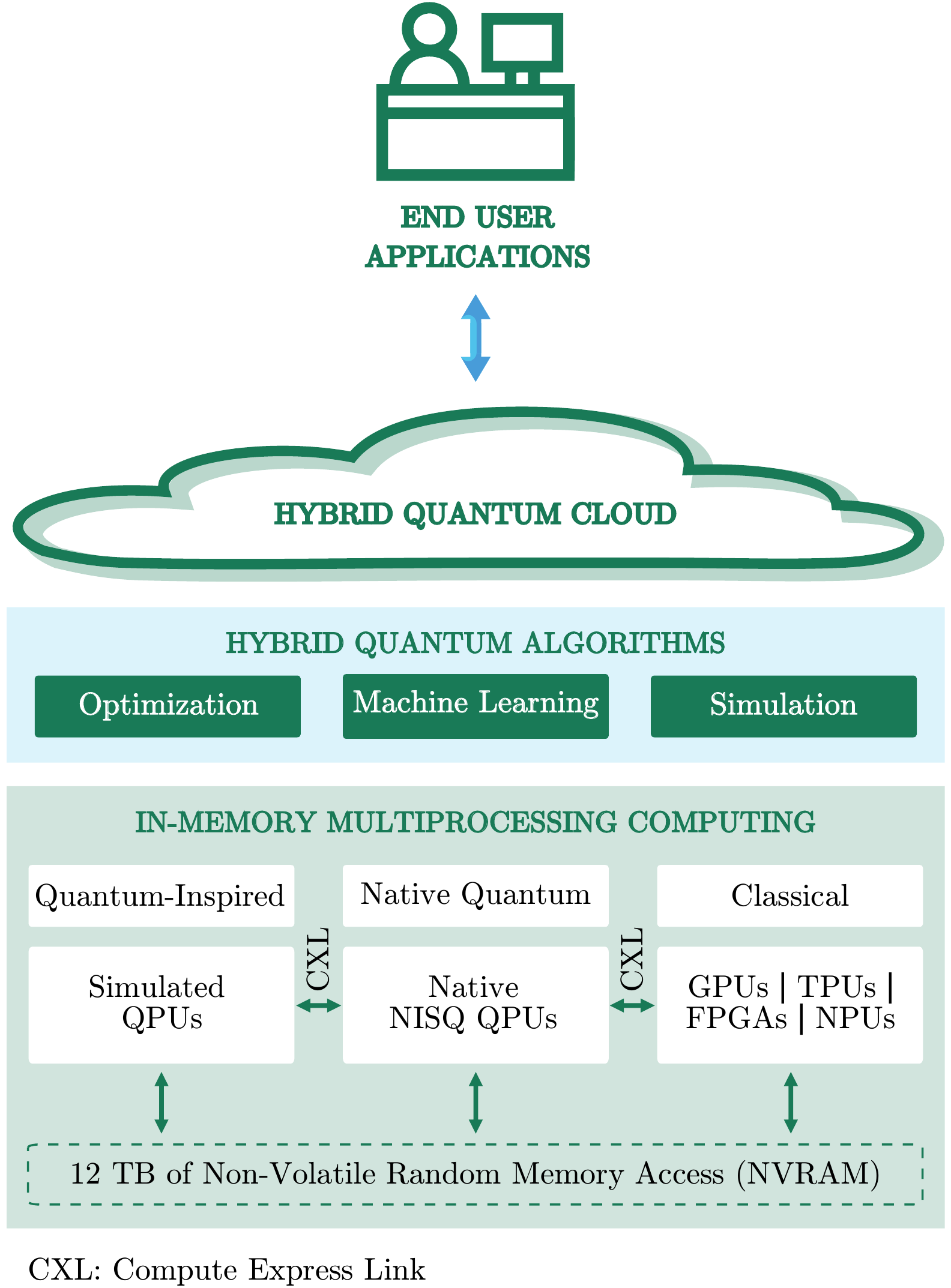}
    \caption{QMware hybrid quantum cloud: accessing applications with quantum algorithms capabilities. Multiprocessing and in-memory compute allows for application-specific practical advantage. The integrated simulated QPUs are able to simulate up to $40$ fault-tolerant qubit circuits. Native QPUs of different topologies and emulations of them are supported by our SDK, but are currently not yet integrated in the hybrid cloud.}
    \label{fig:HCloud2}
\end{figure}

Incorporating these directions addresses scalability, latency and memory-related challenges limiting overall performance today. The QMware cloud drives progress in these three areas and provides a system for disruptive application development around the most demanding computational problems in optimization, machine learning, and simulation, see Fig.~\ref{fig:HCloud2}. 

\subsection{Hybrid Quantum Architecture}
    \tikz [remember picture, overlay] %
    \node [shift={(-1.7cm,0.1cm)}] at (current page.south east) %
    [anchor=south east] %
    {\includegraphics[width=15mm]{figs/Terra-Quantum-Logo-green-e1587399449377.png}};

Quantum circuits are hard to simulate classically as the time and memory needed for simulations and computational cost scales exponentially with the number of qubits~\cite{Preskill2018}. Notably, however, there are classes of algorithms that scale more favourably, although still exponentially, for specific sets of quantum circuits, e.g., tensor networks~\cite{TN2006}. Reports on several high-performance quantum circuit simulators have been published, including full state vector codes built for the CPUs~\cite{Niwa2002, Guerreschi2020}, and/or GPUs, and those that use a mix of algorithm types~\cite{Jones2019}.

Our hybrid computing hardware is based on a unique memory-centric architecture, shown in Fig.~\ref{fig:HCloud2}, whereby the same main memory is accessible by all the different processing units harnessed for problem-solving, including CPUs, GPUs, QPUs, and the like, in a similar manner and to its whole extent. This uniform computing model tackles the challenges around effectively storing and processing both quantum and classical information in one system through implementing a memory bus system for synchronisation between the physical processing units and the main memory. The in-memory processor virtualization enables the realization of simulators, emulators and physical hardware level integrations with native QPUs of all topologies from across the ecosystem in a highly efficient manner. The logical unification through an intermediate representation prevents time consuming copying of information in between process cycles. Through a system that unifies hybrid multi-processing compute, QMware enables the processor units to independently access the same central main memory. It also facilitates the scalable heterogeneous processing which is highly desirable for specific applications, e.g., for image classification using GPUs and QPUs through realizing the Compute Express Link (CXL) between processing units to scale performance.

While many hybrid algorithms feature a continuous loop of classical-quantum-classical interactions as the most time-consuming subroutine, our approach optimizes the quantum-classical interface to provide speed-ups through a hybrid quantum computing pipeline. Furthermore, due to the algorithmic universality and large shared memory capabilities (12 TB per node, next version 32 TB), any quantum circuit built on our hybrid quantum cloud with virtualized qubits can also be run on upcoming native QPUs from across the ecosystem. The other way around, the QMware cloud is capable to import and run any outside developed algorithm. Additionally, the intermediate representation is downwards compatible for future hardware. This gives end users the ability to build modular hybrid quantum algorithms today while maintaining hardware flexibility in the long term.

\subsection{Introducing Quantum into Classical}
    \tikz [remember picture, overlay] %
    \node [shift={(-1.7cm,0.1cm)}] at (current page.south east) %
    [anchor=south east] %
    {\includegraphics[width=15mm]{figs/Terra-Quantum-Logo-green-e1587399449377.png}};

While many use-cases can be tackled using advanced quantum inspired computing, the exponential growth of the required classical resources limits the size of the simulated quantum computational space.

In the case of optimization problems, a quantum annealing~\cite{Salvador2018} or a quantum approximate optimization algorithm (QAOA)~\cite{QAOA}, executed on a 40-qubit device, can handle a problem with 40 binary variables. Partitioning larger problems to be handled by these approaches significantly compromises solution quality. However, using more sophisticated encoding in the hybrid quantum algorithm allows for tackling problems with many thousands of variables, without compromising the solution quality. 

Further extension of the problem complexity requires the native QPUs of the size and quality that are expected to appear on the market in forthcoming years. Many machine learning models benefit from the exponentially larger parameter space while keeping learning simple, e.g., kernel methods~\cite{Kernel2014}. Enlarging the model using native quantum devices allows for more complex data analysis~\cite{ML2021}.

Quantum advantage has been demonstrated on QPUs programmed to execute random instructions that were used to mimic a quantum algorithm\,\cite{Arute2019}. In the previous case, the QPU took 200 seconds to sample a quantum circuit a million times, while the execution time of the same task on a classical supercomputer is several orders of magnitude higher. Recently, a similar experiment was repeated utilizing more qubits and more complex circuits \cite{SupremacyChina2021a, SupremacyChina2021b}. Besides, Ref.~\onlinecite{Perelshtein2020} has implemented a quantum algorithm for a linear system solution that can be used for demonstrating an advantage of quantum phase estimation procedure, which we also use in a quantum sensing protocol~\cite{LAMA}. Such a dramatic speed-up certainly counts as experimental evidence that for specific use cases hybrid quantum computers will surpass purely classical computers that reigned in the past.

{\color{TQgreen2}
\section{Application-Specific Hybrid Quantum Advantage}}

Here, we address three industry-valuable domains, namely optimization, machine learning and simulation that posit heavy challenges and demonstrate how hybrid quantum computing can improve solution performance. 
These challenges and limitations are faced by classical algorithms and existing hardware across these three domains. 
For instance, the inability to find the global minima in feasible time for optimization challenges~\cite{cheeseman1991really}, the large energy intensiveness when training deep learning models~\cite{strubell2019energy} and the inability to simulate large scale complex systems~\cite{georgescu2014quantum}.\\\medskip

\subsection{Optimization}
    \tikz [remember picture, overlay] %
    \node [shift={(-1.7cm,0.1cm)}] at (current page.south east) %
    [anchor=south east] %
    {\includegraphics[width=15mm]{figs/Terra-Quantum-Logo-green-e1587399449377.png}};

The problem of multiparameter optimization in the presence of multiple constraints arises in various subroutines of large-scale business management, such as an optimal resource allocation for improving efficiency, reducing risks and costs, and increasing profit. 
Many {\it discrete} optimization problems are NP-hard~\cite{Karp1972}, rendering the creation of efficient methods for finding the optimal solution impossible on large enough scales. 
The computational challenges that appear when solving such problems include, among others, exponential increases in computational cost with increasing dimensionality and the number of local minima~\cite{cheeseman1991really}.

One of the most advanced classical software approaches to performing large-scale discrete optimizations are combinations of Simplex, the interior point-based methods, Branch and Bound algorithms, mixed-integer linear and quadratic programming, and mixed-integer constrained programming~\cite{1966, Dongarra2000}. These methods are implemented in commercially available solvers, such as CPLEX by IBM~\cite{cplex2009v12} or Gurobi Optimizer by Gurobi~\cite{gurobi}. 
Many important problems have been solved by utilizing these solvers, but their large-scale performance is limited due to the likely exponential increase in optimization complexity\,\cite{Knuth1974}. Even without the exponential increase in complexity, the proper tuning of optimizers becomes unfeasible on a larger scale since the landscape of the cost function becomes much too complex. The latter is instrumental in solving real-life problems, such as applications in industry, where time is often a constraint and compute resources are scarce. Quantum algorithms can help in finding solutions in a faster and more accurate manner~\cite{farhi2001quantum, QAOA, moll2018quantum, PhysRevResearch.2.023302}.\\

Indeed, in the quantum computing community, discrete optimization is considered to be one of the leading candidates to demonstrate a quantum advantage in NISQ devices. Quantum annealers, in contrast to universal quantum computers, are special-purpose hardware developed explicitly for solving optimization problems and sampling tasks~\cite{Salvador2018}. De facto, quantum annealing is an algorithm for solving Ising spin glasses~\cite{2005} inspired by classical simulated annealing, and quantum annealing systems implement this algorithm in quantum hardware.
Such machines implement a noisy version of the Quantum Adiabatic Algorithm, the mathematical equivalent used for formulating the problem mostly being the Quadratic Unconstrained Binary Optimization (QUBO).
Solving a QUBO minimizes a polynomial function of binary variables, with a degree at most two. The QUBO model is an NP-hard discrete optimization problem that lies in a cost function minimization $\min(C)=\min_{\vec{x}}(\vec{x}^TQ\vec{x})$, where $x_i \in \{0,1\}$ is the component of a vector $\vec{x}$ of $n_c$ binary variables and $Q$ is a real and upper triangular matrix. Many known binary-constrained problems can be reduced to QUBO using penalties in the cost function. Remarkably, it was shown in Ref.~\cite{Lucas2014} that many NP-complete and NP-hard problems, including all of Karp’s 21 NP-complete problems, can be reduced to QUBO in a polynomial time.

Among the wide range of the QUBO problems, the MaxCut on an arbitrary graph~\cite{Ausiello1999} is often used to analyse the performance of quantum algorithms. The MaxCut problem is a search of the partition of the graph's nodes into two complementary sets, such that the sum of the weighted edges between these two sets is as large as possible. The MaxCut problem is related to logistics and planning, such as machine scheduling, traffic message management, computer-aided design, image recognition, and unsupervised machine learning problems, such as clustering and various financial optimizations. The MaxCut problem can be formulated as the QUBO problem, which lies in a minimization of the following quadratic function:
\begin{equation}
    E = -\sum_{i,j=1}^{n_c}d_{ij}(x_i-x_j)^2,
    \label{energy}
\end{equation}
where $d_{ij}$ is the weight of the edge between $i$-th and $j$-th nodes in the studied graph. The solution is a binary string $\vec{x}$ of nodes' indicators that show the correspondence to one of two sets. The elements of the QUBO matrix, in turn, are $Q_{ij}=2d_{ij} (i > j)$ and $Q_{ii}=-\sum_j d_{ij}$.

Inspired by quantum annealing, the optical Coherent Ising Machine (CIM) was developed to find solutions using light pulses in a hybrid electro-optical loop~\cite{CIM2019}. One of the most significant advantages of such a machine is the effective full connectivity that encodes the whole problem without sacrificing the majority of qubits to overcome the QPU connectivity issue. For instance, Ref.~\cite{Hamerly2019} presents a comparison between D-Wave and CIM in the MaxCut problem solution: the significant time-to-solution difference for graphs with over 50 nodes (up to 200 nodes) was observed for dense problems where CIM demonstrates better results. The difference in performance between the sparsely connected D-Wave machine and the fully connected CIMs provides strong experimental support for increasing the qubit-connectivity on quantum annealers. In general, CIM offers intriguing and prospective platforms for studying discrete optimization problems powered by the speed of light and effective electrical feedback. However, implementing a CIM comes with formidable engineering challenges~\cite{CIM2017}. 

\begin{figure*}[ht]
    \includegraphics[width=1\linewidth]{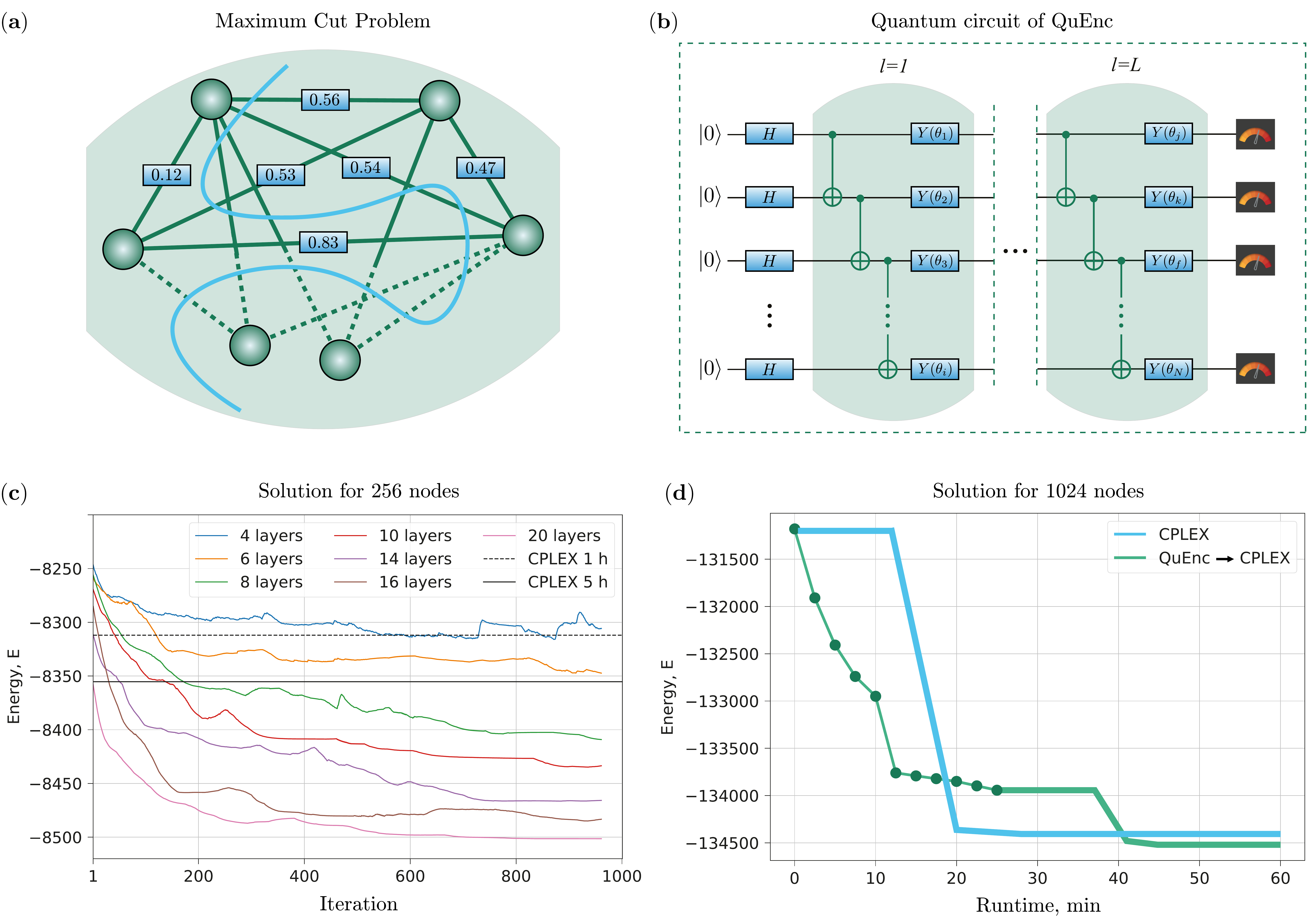}
    \caption{
    (a) The schematic drawing of the weighed fully connected graph. 
    The cut that gives the largest value is called Maximum Cut (orange line) and the search of such a cut is an NP-hard problem.
    (b) The quantum circuit used in the QuEnc algorithm with fully entangled quantum state. The circuits parameterised are optimized to sample the classical solution with the lowest energy (cost) value.
    (c) The QuEnc and CPLEX performance on complete graphs with 256 nodes.
    Our quantum algorithm with $\geq 20$ layers finds significantly more accurate solution with lower energy in 1 minute than the CPLEX in 5 hours.
    (d) The cost function obtained via quantum algorithm and high-performance CPLEX on complete weighted graphs with 1024 nodes as function of runtime. The solution obtained by simulating quantum circuit with $50$ layers in 25 minutes and utilization the resulting solution as an initial point for the CPLEX solver, which improves the obtained solution for 35 minutes. Such a hybrid pipeline finds more accurate solution with lower energy in 1 hour than the pure CPLEX.
    } 
    \label{MaxCut}
\end{figure*}

Besides specially designed hardware, one of the most promising approaches in tackling discrete optimization on universal quantum devices is variational quantum algorithms. In such algorithms, a parameterised quantum network is iteratively optimized using classical computing. A paradigm of variational circuits is close to the neural networks approach, where a deep network captures the fundamental features of the problem. For several decades, research in machine learning was focused on models that can provide theoretical guarantees for their performance. However, in recent years, methods based on heuristics have become dominant, especially for deep models, partly due to an abundance of data and computational resources. Similarly, a formal proof of quantum advantages of variational quantum algorithms was not found yet, but applications utilizing NISQ-devices to solve real-world problems using such algorithms are already being explored~\cite{VQC2021}.

For instance, algorithms such as QAOA have been applied to solve NP-hard QUBO problems. However, recent experiments have highlighted the challenges in implementing the QAOA on problem graphs that diﬀer from the native hardware topology, even for small system sizes. The latest solution of the MaxCut on a dense problem processes a 24-node graph using QAOA~\cite{Harrigan2021}. There are other emerging quantum approaches such as the Filtering Variational Quantum Eigensolver (F-VQE) for combinatorial optimization problems that have proved to be more performant than the original Variational Quantum Eigensolver (VQE) algorithm and QAOA~\cite{amaro2021filtering}.
    \tikz [remember picture, overlay] %
    \node [shift={(-1.7cm,0.1cm)}] at (current page.south east) %
    [anchor=south east] %
    {\includegraphics[width=15mm]{figs/Terra-Quantum-Logo-green-e1587399449377.png}};

Most solutions based on modern variational algorithms, e.g., QAOA, or quantum annealing, require a significant number of qubits to solve real-world MaxCut related problems. Not considering the chip topology and qubit interconnectivity, the number of qubits necessary is equal to the number of classical variables. Such an encoding requires quantum resources that current NISQ devices can't provide due to a limited number of qubits and limited qubit-interconnectivity.\\

Therefore, we explore novel encoding and optimization techniques for variational algorithms in a gate-based computing framework. We utilize the hardware-efficient ansatz \cite{VQC2021}, and amplitude encoding scheme \cite{Tan2021} to transfer the classical optimization problem into the optimal quantum state search. The algorithm is called {\it QuEnc} because of the algorithm's underlying {\it Qu}antum {\it Enc}oding method. Inspired by available quantum machine learning tools \cite{Broughton2020TensorFlowQA}, we learn the circuit parameters that provide the desired quantum state, corresponding to the optimal classical solution. The expressibility of the quantum circuit is set by the number of layers that control the number of circuit parameters. Here, we do not need to perform the full state tomography since we operate only the Z-projection of the state. The whole scheme is described in detail in Ref.\,\cite{QuEnc_patent}.

We apply QuEnc to the fully connected MaxCut problem and, leveraging the amplitude encoding, we solve MaxCut with hundreds and thousands of nodes -- a much larger scale than has previously been possible with quantum annealing or QAOA. Since QuEnc processes much larger problems and, therefore, can not be compared with existing quantum alternatives, we use high-performance mathematical programming solver CPLEX to analyse the performance of the QuEnc. As the solution's performance heavily depends on the classical hardware, we consider two cases: (i) the average local computing setup with 32\,GB of RAM and 6 CPU cores, and (ii) the advanced hardware provided by QMware with 100 CPU cores and 12\,TB of RAM. In both cases, the quantum algorithm was implemented and simulated using the <basiq> Python SDK with highly efficient C++ kernels~\cite{QMWare}.

To define a MaxCut problem, we create a random weighted graph with weights of edges laying in $[0.01, 1]$, and minimize the energy function from Eq.\,\ref{energy}. Schematic drawing of the graph is shown in Fig.~\ref{MaxCut}(a).

In case (i), we focus on 256-node fully connected graph and compare CPLEX and simulated QuEnc approaches. We need to bear in mind that the exact solution of such a complex problem is impossible since we face $2^{256}$ possible solutions. We apply QuEnc algorithm with the fully entangled circuit presented in Fig.\,\ref{MaxCut}(b). Due to the amplitude encoding QuEnc provides a great advantage to solving larger problems operating as an encoder -- we reduce the discrete $n_c$-parameter optimization to continuous O($\log{n_c}$)-parameter problem. Its convergence, reduction of the energy (cost) as a function of learning iterations, is presented in Fig.\,\ref{MaxCut}(c). Different colors correspond to different numbers of layers -- the increase in layers leads to an increase in the number of optimized parameters. The simulated 20-layer QuEnc finds the solution with cost $-8,500$ in 30 minutes, while CPLEX finds the cost $-8,360$ in 5 hours, indicating superior performance over the hybrid QuEnc. The transfer of the simulated QuEnc to real QPU provides a further increase in speed (2-3x for that problem considering superconducting QPU). While the presence of noise is expected to limit the accuracy of the algorithm, remarkably, arising errors may even help to avoid local minima during the convergence that only benefits the accuracy. The comprehensive study of such an important issue is a subject of further work.

In case (ii), CPLEX as a well-tuned and flexible solver leveraging the full power of QMware can compete with QuEnc and finds the solution with even better cost value. Therefore, we are able to consider a larger scale problem with a 1024-node graph, where the optimization landscape is much more complex.
Using encoding techniques helped us to find the solution with good cost value very fast, but usually that solution could be improved upon even more.
Here, to leverage the whole power of QuEnc we introduce the hybrid pipeline with the high-performance classical solver. Mainly, we presolve the problem using QuEnc obtaining the solution with good cost and then use that solution as the initial point for the CPLEX solver.

The QuEnc and CPLEX convergence, reduction of the energy (cost) as function of execution time, is presented in Fig.~\ref{MaxCut}(d). Green dots show the QuEnc solution convergence, which reach $-13,400$ in 25 minutes. The green line is the convergence of the CPLEX that starts from the point that was found by QuEnc. The plato arises due to the CPLEX internal processes, such as tree building. The blue line corresponds to the pure CPLEX solution started at a random point, whose cost is close to the initial cost of QuEnc that also started from a random point. The QuEnc$\rightarrow$CPLEX pipeline finds better cost of $-134,520$ than CPLEX, $-134,406$, providing 0.085\% improvement. We expect that the improvement would be more prominent with the increase in the problem size and enchanting QuEnc that falls within the scope of future work.

\subsection{Machine Learning}
    \tikz [remember picture, overlay] %
    \node [shift={(-1.7cm,0.1cm)}] at (current page.south east) %
    [anchor=south east] %
    {\includegraphics[width=15mm]{figs/Terra-Quantum-Logo-green-e1587399449377.png}};

Quantum machine learning, both improving quantum technologies with artificial intelligence and enhancing classical machine learning utilizing quantum effects, vastly illustrates the power of hybrid quantum computing. 

First, classical machine learning is becoming increasingly more important for a variety of tasks in quantum information technologies~\cite{dunjko2018machine,carleo2019machine,biamonte2017quantum}. For instance, reinforcement learning agents are being used for control~\cite{fosel2018reinforcement,bukov2018reinforcement,xu2019generalizable}, error-correction~\cite{nautrup2019optimizing,sweke2020reinforcement}, and designing new quantum protocols and experiments~\cite{melnikov2018active,wallnofer2020machine,melnikov2020setting}. The latter is motivated by the unknown reachability of various configurations in quantum experiments~\cite{krenn2016automated,krenn2020computer}. These works show that machine learning can offer dramatic advances in how complicated experiments are generated. In addition to reinforcement learning, supervised learning systems were found useful for, e.g., studying quantum advantage over a classical approach~\cite{melnikov2019predicting,melnikov2020machinetransfer,moussa2020quantum}, reconstructing quantum states of physical systems~\cite{shang2019reconstruction,torlai2019integrating,palmieri2020experimental,ding2020retrieving}, and learning compact representations of these states~\cite{carleo2017solving,gao2017efficient}. These studies have revealed that deep learning networks can identify complex patterns and trends in data. It would not be possible without powerful computers and special-purpose hardware capable of implementing deep networks with billions of parameters~\cite{Le2012}. If machine learning implemented on HPC could substantially improve quantum devices, the potential impact would be tremendous.

Second, quantum technologies can massively assist the most advanced machine learning frameworks -- highly autonomous systems that outperform humans at most economically valuable work require considerable computational resources, limiting their performance. Quantum computing models can potentially improve the training process of existing classical models~\cite{Neven2012QBoostLS, PhysRevLett.113.130503, dunjko2018machine, saggio2021experimental}, which allows for finding better extreme points in an objective function landscape or the same optima with fewer iterations. These methods allow for polynomial speedups, which are crucial for large and complex problems, where minor improvements give noticeable gains. Besides, the recent experiments show that quantum models can sample intricate probability distributions in a polynomial time~\cite{lund2017quantum}, while the same classical sampling could be exponentially difficult. Among many other methods, the most promising are quantum neural networks \cite{Farhi, PhysRevA.98.042308, mcclean2018barren, beer2020training, Broughton2020TensorFlowQA} and quantum kernels \cite{Havlek2019} that are expected to beat classical models with current noisy quantum devices.

\begin{figure*}[ht]
    \includegraphics[width=1\linewidth]{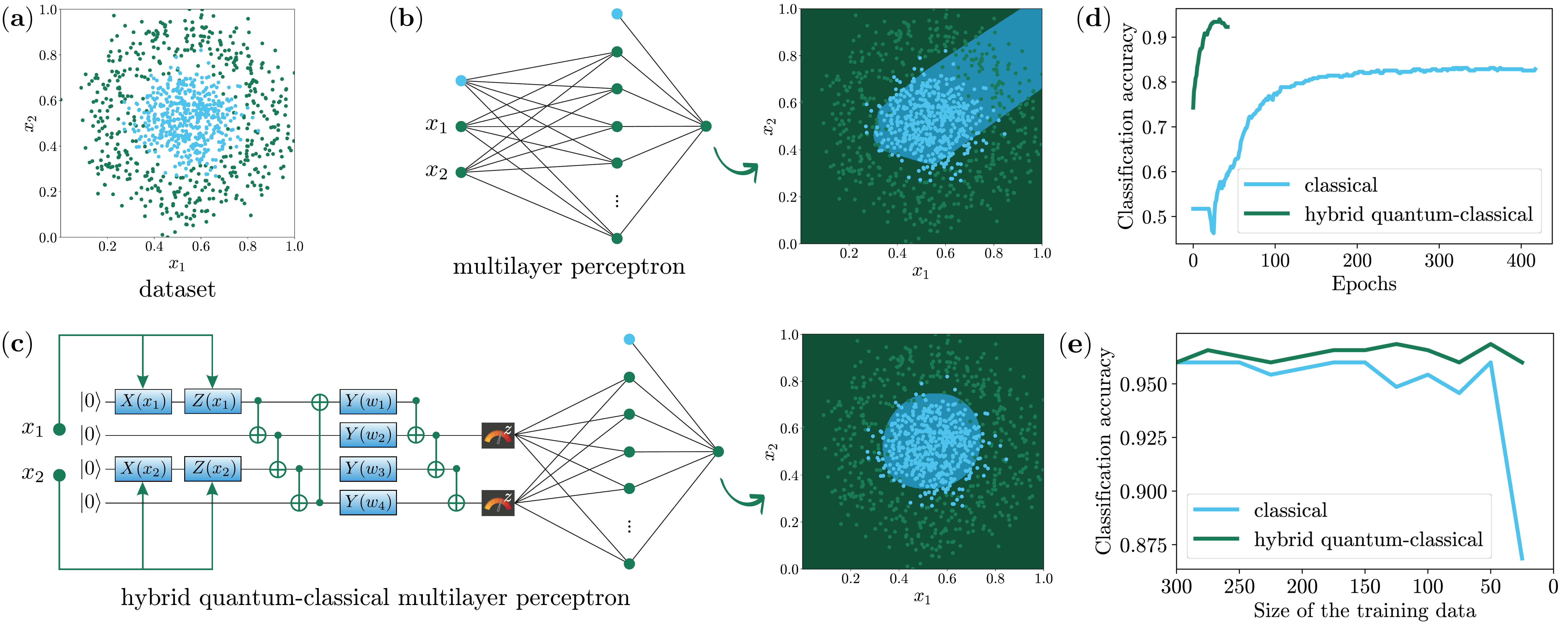}
    \caption{(a) Circles dataset from Scikit~\cite{CirclesDataset}. 1000 points are generated: 300 for training procedure and 700 for evaluating model's performance. (b) Classic neural network's architecture. A classification function learned by this network is a dependence of the output class on the coordinates $x_1$ and $x_2$. (c) Hybrid quantum neural network's architecture. A classification function learned by this network is a dependence of the output class on the coordinates $x_1$ and $x_2$. (d) Accuracy on test data for each epoch during the training procedure. Results are averaged over 10 independent models. (e) Dependence of the classification accuracy on the training data size.}
    \label{fig_classification}
\end{figure*}

Since NISQ devices limit the freedom in the machine learning model choice, we focus on the hybrid pipelines for classification and regression as the most suitable approaches. Hybrid quantum-classical solvers implemented in the hybrid quantum cloud can provide higher efficiency in training by requiring lesser iterations, and can show a higher prediction accuracy as we demonstrate next.

\subsubsection{Hybrid Machine Learning Advantages in Classification}
    \tikz [remember picture, overlay] %
    \node [shift={(-1.7cm,0.1cm)}] at (current page.south east) %
    [anchor=south east] %
    {\includegraphics[width=15mm]{figs/Terra-Quantum-Logo-green-e1587399449377.png}};

To illustrate the difference that quantum circuits introduce in machine learning, we first consider a standard benchmark Scikit dataset~\cite{CirclesDataset} used to test small-scale classification algorithms. The dataset can be visualized as a large circle containing a smaller circle in 2D, see Fig.~\ref{fig_classification}(a).

As a classical machine learning solution to this binary classification problem, we use a multilayer perceptron model with 3 neural network layers, shown in Fig.~\ref{fig_classification}(b). The first layer consists of 2 input neurons and a bias neuron, followed by a hidden layer with 40 neurons and a bias, and a single output neuron. Neural network layers are fully connected, leading to 161 weights in the network.

As a hybrid quantum machine learning solution to this binary classification problem, we use a hybrid quantum-classical multilayer perceptron. This hybrid quantum neural network consists of both quantum and classical neural network parts: a 4-qubit quantum circuit followed by 3 layers of neurons for a classical part, shown in Fig.~\ref{fig_classification}(c). The classical part of the hybrid model is the same as of a classical multilayer perceptron, but without a bias neuron. The quantum circuit has 4 variational parameters, leading to a total of 125 weights in the hybrid quantum neural network (HQNN).

On Fig.~\ref{fig_classification}(d) one can see the training procedure of the best model over hundred independent models. Both models have been trained with stochastic gradient descent, Adam optimizer with learning rate $1 \times 10^{-2}$, and the Binary cross entropy loss function. The HQNN algorithm achieves a 13\% higher accuracy on the test data (0.831 and 0.940 accuracy for the classical and the hybrid model, respectively), but also converges much faster than the classical counterpart (317 and 32 epochs for the classical and the hybrid model, respectively). On Figs.~\ref{fig_classification}(b)-(c) one can additionally see a qualitative difference in how well both models separated data points belonging to different classes with the training size of 25 samples.

Moreover, by reducing the size of the training data, the classical model's ability to learn decreases substantially. Meanwhile, the HQNN does not show any difference and demonstrates accuracy above 90\%, as can be seen in Fig.~\ref{fig_classification}(e). The obtained results are significant, since the most difficult part of commercial machine learning tasks is collecting data and labelling it. Therefore, it is very useful that this hybrid model manages to learn well even on small datasets.

The HQNN model used to solve the binary classification problem, can be used for a continuous variable output, namely it can deal with regression problems as we show next.

\subsubsection{Hybrid Machine Learning Advantages in Regression}
    \tikz [remember picture, overlay] %
    \node [shift={(-1.7cm,0.1cm)}] at (current page.south east) %
    [anchor=south east] %
    {\includegraphics[width=15mm]{figs/Terra-Quantum-Logo-green-e1587399449377.png}};

The regression problem under consideration is represented by the Boston housing dataset~\cite{BostonDataset}. This dataset is used as a comparison of machine learning models, testing of various algorithms and based on information gathered by the US Census Service regarding housing in the Massachusetts capital, Boston. The dataset contains 506 samples and 13 feature variables such as average number of rooms per dwelling, pupil-teacher ratio, and per capita crime rate. Our goal is to determine the value of the median price of owner-occupied homes based on these features. The data was originally published in~\cite{harrison1978hedonic}. The title of the article suggests that people are willing to pay more for clean air and that the price of houses depends on the surrounding area. It is worth noting that house prices do not exceed 50,000\$. This is due to the fact that the Census Service censored the data and put a price cap of 50,000\$. After all these years, it is impossible to reliably know the real prices of houses.

\begin{figure*}[ht]
    \includegraphics[width=1\linewidth]{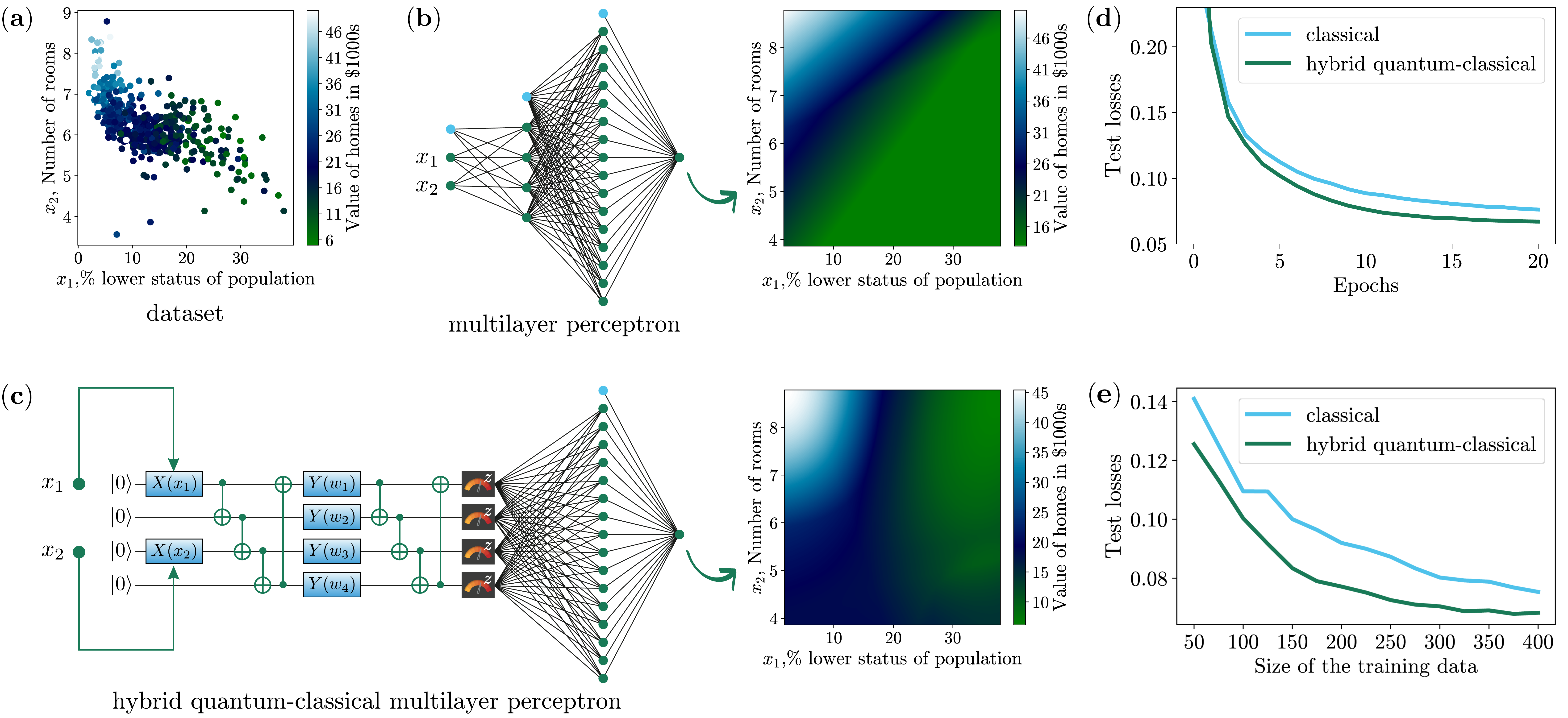}
    \caption{(a) Boston Housing dataset from Scikit~\cite{BostonDataset}. (b) Classic neural network's architecture. A function learned by this network is a dependence of the value of houses on the number of rooms and the \% of lower status of population. (c) Hybrid quantum neural network's architecture. A function learned by this network is a dependence of the value of houses on the number of rooms and the \% of lower status of population. (d) Loss on the test data for epoch during the training procedure. Results are averaged over 100 independent models. (e) Dependence of the test loss on the training data size.}
    \label{fig_regression}
\end{figure*}

The data was downloaded using the Scikit-learn library~\cite{BostonDataset}. After analysing the data, we find a relationship between the target variable with other features and selected two features: number of rooms (average number of rooms per dwelling) and LSTAT (percentage of lower status of the population). We built a machine learning model for determining the prices of Boston houses based on two features we selected. Next, we split the dataset into training and test samples, 80\% and 20\%, respectively.

Similar to the classification task, we use the multilayer perceptron for solving the regression problem. The architecture of our classic network is shown in Fig.~\ref{fig_regression}(b) and consists of three linear layers with the ReLU activation functions. As for the hybrid neural network, we construct it by replacing the first classical fully-connected layer with a quantum variational layer with four qubits, as shown in Fig.~\ref{fig_regression}(c). Since we decided to use only two features, the dimension of the input data is equal to two. Therefore, we encoded the input information into the angles of rotations along the $x$-axis on the first and third qubits. It should be noted that the number of parameters in our quantum variation circuit is four versus twelve parameters in the first classical layer in the classical analogue of our hybrid network.

We train our classical and hybrid models using stochastic gradient descent, Adam optimizer with the learning rate $3 \times 10^{-3}$ and compute the mean squared error loss function. Mean absolute error is used to evaluate the model's performance on the test set. To compare the classical and hybrid neural networks, in Fig.~\ref{fig_regression}(d) we plot depicting the dependence of test losses on the number of epochs during the training procedure. This plot was obtained by averaging the results across one hundred independent models. The HQNN has an advantage over the classical neural network, as the test loss is 12\% lower in the hybrid case (0.076 and 0.067 loss values for the classical and hybrid models, respectively). 

Another interesting feature is scaling. In Fig.~\ref{fig_regression}(e) we show test loss as a function of the number of training data samples used for training. Every point of this plot, taken every 50 steps in training data size, is obtained by averaging across 100 independent models. Similar to the classification problem, in the regression problem we observe that the hybrid quantum advantage is robust, and the quantum advantage is observable in an entire range of training data sizes. The advantage that we observe ranges between 12\% improvement for the training set size of 400 (see Fig.~\ref{fig_regression}(d)) and 16\% improvement for the training set size of 125 (0.109 and 0.092 loss values for the classical and the hybrid model, respectively). Similar to the Circles Scikit dataset, we observe a better advantage for smaller dataset sizes.

\subsection{Quantum-inspired simulation}
    \tikz [remember picture, overlay] %
    \node [shift={(-1.7cm,0.1cm)}] at (current page.south east) %
    [anchor=south east] %
    {\includegraphics[width=15mm]{figs/Terra-Quantum-Logo-green-e1587399449377.png}};

The tantalizing goal of quantum computing is to perform calculations beyond the reach of any classical computer, e.g., high-performance modelling of complex physical and biological systems. The NISQ devices suffer from many sources of errors, which limit the degree of entanglement and their current performance. However, quantum-enabled algorithms, even being implemented via a tensor network on classical hardware, can provide an advantage at solving certain problems.

\begin{figure}[h!]
    \centering
    \includegraphics[width=1\linewidth]{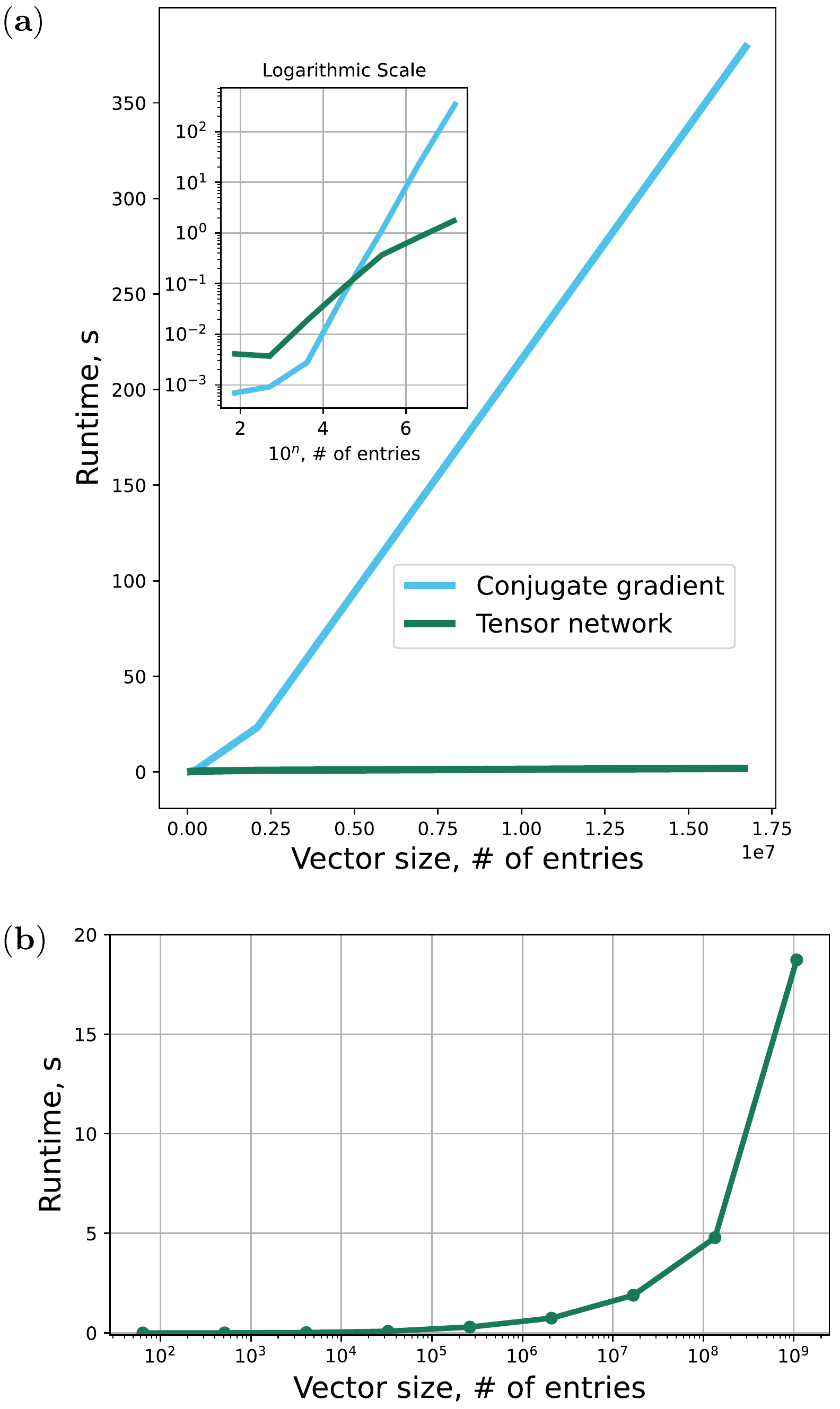}
    \caption{(a) The runtime of the Poisson equation solution as a function of the space discretisation for the conjugate gradient method (black) and tensor networks (red). The embedding shows the log-log dependency. Tensor networks provide an exponential advantage in the solution of such a problem. (b) The runtime of the Poisson equation solution as a function of the space discretization for the tensor network-based method. Such a method allows solving problem with $10^9$ points in space discretisation in less than 20 seconds using just two CPUs.}
    \label{Fig:Simulation}
\end{figure}

In general, tensor decompositions and tensor networks, which were initially introduced in quantum physics for multiparticle system analysis~\cite{DMRG}, are emerging as promising methods for high-dimensional problems, simulation in particular. The main advantage of tensor networks is logarithmic scaling in the studied problem's dimension for some tasks, which is similar to scaling expected in quantum computers. Moreover, the topology of tensor networks is similar to quantum circuits architecture, which makes them an efficient tool for low-entangled quantum computer virtualization and simulation of certain physical systems~\cite{Zhou2020}.

Many physical and biological simulations depend on solving partial differential equations describing the physical processes behind studied systems. Using tensor decomposition, one can find an efficient way to solve differential equations saving memory and speeding up the solution.

Here, as an illustrative example, we study the solution of second-order linear partial differential equations via tensor networks. Mainly, we consider the Poisson equation, which is a generalization of Laplace's equation, that is frequently used in various areas of science and engineering, e.g., computational fluid dynamics~\cite{Batchelor2000}, electrostatics~\cite{Griffiths1999}, the theory of Markov chains~\cite{Meyn2009}, and density functional theory and electronic structure calculations~\cite{Engel2011}.

We solve the Poisson equation using tensor networks by discretisation on a regular grid in Cartesian coordinates and processing the resulting system of linear equations. Leveraging the fact that the matrix corresponding to the second-order derivative can be represented as a matrix product operator~\cite{Kazeev2012}, we can solve such a system of linear equations employing a method based on an alternating minimal energy solver~\cite{AMEN}. Those types of algorithms provide polylogarithmic scaling in runtime and memory, offering exponential speedup in comparison with conventional classical methods that have polynomial scaling.

Let us consider the solution of the Poisson equation in a 3-dimensional space with zero boundary condition
\[ -\Delta u = 1 \,\,\,\text{in} \,\, \Omega = [0, 1]^3, \,\, u \mid_{\partial \Omega} \,\, = 0. \]
We reduce the differential equation to a linear system, represent it as a tensor network, process and contract tensors to obtain a classical solution. In order to verify the solution we consider one of the most powerful tools in a linear systems solver -- a conjugate gradient method. The runtime of the solution for both a tensor network and conjugate gradient methods as a function of discretization accuracy, number of grid points, is presented in Fig.\,\ref{Fig:Simulation}(a). It is clear that the tensor network solver provides an exponentially faster solution in comparison with classical approaches with similar scaling as proposed quantum algorithms~\cite{Wang2020}. However, in comparison with NISQ-device implementation, our method allows us to solve large-scale problems in seconds, as presented in Fig.\,\ref{Fig:Simulation}(b). Here, we solve the Poisson equation on $10^9$ spatial points with perfect fidelity via tensor networks in less than 20 seconds using just two Intel Xeon 2.2\,GHz CPUs and 12 Gb of RAM.

Leveraging the full power of the QMware hybrid quantum cloud, a tensor network approach can handle larger and more complex problems that will be studied in the future. Such a method can be extended to more complex differential equations in many areas, including computational fluid dynamics, e.g., Boltzmann equation, financial simulations, e.g., Fokker-Planck equation, aerodynamics and heat transfer, e.g., parabolic differential equations.\\

{\color{TQgreen2}
\section{Conclusion}}
    \tikz [remember picture, overlay] %
    \node [shift={(-1.7cm,0.1cm)}] at (current page.south east) %
    [anchor=south east] %
    {\includegraphics[width=15mm]{figs/Terra-Quantum-Logo-green-e1587399449377.png}};

This paper presents a hybrid quantum architecture for hardware and software targeting maximizing application-specific practical advantage today. This architecture puts forth hybrid quantum algorithms for optimization, machine learning and simulation. We demonstrate solution performance for problems across these application domains, through examples of discrete graph-based optimization, classification, regression and partial differential equations. In the MaxCut optimization problem with 256 nodes with the limited hardware, the QuEnc algorithm provided a 1.7\% better solution in 1 minute than the CPLEX solver in 5 hours. The hybrid QuEnc algorithm achieved a 0.085\% better solution on a 1024 node MaxCut problem at the fixed runtime. 
Next, in classification, the hybrid quantum neural network provided, compared to a classical analogue, a 13\% higher accuracy and a much better convergence time (32 epochs instead of 317) on the Circles dataset. In regression, the hybrid quantum neural network outperformed the classical analogue by 12-16\% depending on the training set size. Finally, in solving the Poisson equation, tensor networks were shown to be exponentially faster than the conjugate gradient method providing an improvement in a runtime by several orders of magnitude.

Through our hybrid quantum cloud, QMware, we hence make these demonstrated benefits accessible in an industrial context. The introduced hybrid quantum-classical approach looks to meet the need of industrial users to get the best compute results for their applications, irrespective of whether the underlying hardware is classical or quantum. This paves the way toward accelerating quantum adoption for the benefit of business and society.

\bibliography{aipsamp}

\providecommand{\noopsort}[1]{}\providecommand{\singleletter}[1]{#1}%
\begin{thebibliography}{99}%
\makeatletter
\providecommand \@ifxundefined [1]{%
 \@ifx{#1\undefined}
}%
\providecommand \@ifnum [1]{%
 \ifnum #1\expandafter \@firstoftwo
 \else \expandafter \@secondoftwo
 \fi
}%
\providecommand \@ifx [1]{%
 \ifx #1\expandafter \@firstoftwo
 \else \expandafter \@secondoftwo
 \fi
}%
\providecommand \natexlab [1]{#1}%
\providecommand \enquote  [1]{``#1''}%
\providecommand \bibnamefont  [1]{#1}%
\providecommand \bibfnamefont [1]{#1}%
\providecommand \citenamefont [1]{#1}%
\providecommand \href@noop [0]{\@secondoftwo}%
\providecommand \href [0]{\begingroup \@sanitize@url \@href}%
\providecommand \@href[1]{\@@startlink{#1}\@@href}%
\providecommand \@@href[1]{\endgroup#1\@@endlink}%
\providecommand \@sanitize@url [0]{\catcode `\\12\catcode `\$12\catcode
  `\&12\catcode `\#12\catcode `\^12\catcode `\_12\catcode `\%12\relax}%
\providecommand \@@startlink[1]{}%
\providecommand \@@endlink[0]{}%
\providecommand \url  [0]{\begingroup\@sanitize@url \@url }%
\providecommand \@url [1]{\endgroup\@href {#1}{\urlprefix }}%
\providecommand \urlprefix  [0]{URL }%
\providecommand \Eprint [0]{\href }%
\providecommand \doibase [0]{https://doi.org/}%
\providecommand \selectlanguage [0]{\@gobble}%
\providecommand \bibinfo  [0]{\@secondoftwo}%
\providecommand \bibfield  [0]{\@secondoftwo}%
\providecommand \translation [1]{[#1]}%
\providecommand \BibitemOpen [0]{}%
\providecommand \bibitemStop [0]{}%
\providecommand \bibitemNoStop [0]{.\EOS\space}%
\providecommand \EOS [0]{\spacefactor3000\relax}%
\providecommand \BibitemShut  [1]{\csname bibitem#1\endcsname}%
\let\auto@bib@innerbib\@empty
%</preamble>
\bibitem [{\citenamefont {Preskill}(2018)}]{Preskill2018}%
  \BibitemOpen
  \bibfield  {author} {\bibinfo {author} {\bibfnamefont {J.}~\bibnamefont
  {Preskill}},\ }\bibfield  {title} {\bibinfo {title} {Quantum computing in the
  {NISQ} era and beyond},\ }\href@noop {} {\bibfield  {journal} {\bibinfo
  {journal} {Quantum}\ }\textbf {\bibinfo {volume} {2}},\ \bibinfo {pages} {79}
  (\bibinfo {year} {2018})}\BibitemShut {NoStop}%
\bibitem [{\citenamefont {Bharti}\ \emph {et~al.}(2022)\citenamefont {Bharti},
  \citenamefont {Cervera-Lierta}, \citenamefont {Kyaw}, \citenamefont {Haug},
  \citenamefont {Alperin-Lea}, \citenamefont {Anand}, \citenamefont {Degroote},
  \citenamefont {Heimonen}, \citenamefont {Kottmann}, \citenamefont {Menke}
  \emph {et~al.}}]{bharti2022noisy}%
  \BibitemOpen
  \bibfield  {author} {\bibinfo {author} {\bibfnamefont {K.}~\bibnamefont
  {Bharti}}, \bibinfo {author} {\bibfnamefont {A.}~\bibnamefont
  {Cervera-Lierta}}, \bibinfo {author} {\bibfnamefont {T.~H.}\ \bibnamefont
  {Kyaw}}, \bibinfo {author} {\bibfnamefont {T.}~\bibnamefont {Haug}}, \bibinfo
  {author} {\bibfnamefont {S.}~\bibnamefont {Alperin-Lea}}, \bibinfo {author}
  {\bibfnamefont {A.}~\bibnamefont {Anand}}, \bibinfo {author} {\bibfnamefont
  {M.}~\bibnamefont {Degroote}}, \bibinfo {author} {\bibfnamefont
  {H.}~\bibnamefont {Heimonen}}, \bibinfo {author} {\bibfnamefont {J.~S.}\
  \bibnamefont {Kottmann}}, \bibinfo {author} {\bibfnamefont {T.}~\bibnamefont
  {Menke}}, \emph {et~al.},\ }\bibfield  {title} {\bibinfo {title} {Noisy
  intermediate-scale quantum algorithms},\ }\href@noop {} {\bibfield  {journal}
  {\bibinfo  {journal} {Rev. Mod. Phys.}\ }\textbf {\bibinfo {volume} {94}},\
  \bibinfo {pages} {015004} (\bibinfo {year} {2022})}\BibitemShut {NoStop}%
\bibitem [{\citenamefont {Bravyi}\ \emph {et~al.}(2016)\citenamefont {Bravyi},
  \citenamefont {Smith},\ and\ \citenamefont {Smolin}}]{bravyi2016trading}%
  \BibitemOpen
  \bibfield  {author} {\bibinfo {author} {\bibfnamefont {S.}~\bibnamefont
  {Bravyi}}, \bibinfo {author} {\bibfnamefont {G.}~\bibnamefont {Smith}},\ and\
  \bibinfo {author} {\bibfnamefont {J.~A.}\ \bibnamefont {Smolin}},\ }\bibfield
   {title} {\bibinfo {title} {Trading classical and quantum computational
  resources},\ }\href@noop {} {\bibfield  {journal} {\bibinfo  {journal} {Phys.
  Rev. X}\ }\textbf {\bibinfo {volume} {6}},\ \bibinfo {pages} {021043}
  (\bibinfo {year} {2016})}\BibitemShut {NoStop}%
\bibitem [{\citenamefont {McClean}\ \emph {et~al.}(2016)\citenamefont
  {McClean}, \citenamefont {Romero}, \citenamefont {Babbush},\ and\
  \citenamefont {Aspuru-Guzik}}]{mcclean2016theory}%
  \BibitemOpen
  \bibfield  {author} {\bibinfo {author} {\bibfnamefont {J.~R.}\ \bibnamefont
  {McClean}}, \bibinfo {author} {\bibfnamefont {J.}~\bibnamefont {Romero}},
  \bibinfo {author} {\bibfnamefont {R.}~\bibnamefont {Babbush}},\ and\ \bibinfo
  {author} {\bibfnamefont {A.}~\bibnamefont {Aspuru-Guzik}},\ }\bibfield
  {title} {\bibinfo {title} {The theory of variational hybrid quantum-classical
  algorithms},\ }\href@noop {} {\bibfield  {journal} {\bibinfo  {journal} {New
  J. Phys.}\ }\textbf {\bibinfo {volume} {18}},\ \bibinfo {pages} {023023}
  (\bibinfo {year} {2016})}\BibitemShut {NoStop}%
\bibitem [{\citenamefont {Li}\ \emph {et~al.}(2017)\citenamefont {Li},
  \citenamefont {Yang}, \citenamefont {Peng},\ and\ \citenamefont
  {Sun}}]{li2017hybrid}%
  \BibitemOpen
  \bibfield  {author} {\bibinfo {author} {\bibfnamefont {J.}~\bibnamefont
  {Li}}, \bibinfo {author} {\bibfnamefont {X.}~\bibnamefont {Yang}}, \bibinfo
  {author} {\bibfnamefont {X.}~\bibnamefont {Peng}},\ and\ \bibinfo {author}
  {\bibfnamefont {C.-P.}\ \bibnamefont {Sun}},\ }\bibfield  {title} {\bibinfo
  {title} {Hybrid quantum-classical approach to quantum optimal control},\
  }\href@noop {} {\bibfield  {journal} {\bibinfo  {journal} {Phys. Rev. Lett.}\
  }\textbf {\bibinfo {volume} {118}},\ \bibinfo {pages} {150503} (\bibinfo
  {year} {2017})}\BibitemShut {NoStop}%
\bibitem [{\citenamefont {Zhu}\ \emph {et~al.}(2019)\citenamefont {Zhu},
  \citenamefont {Linke}, \citenamefont {Benedetti}, \citenamefont {Landsman},
  \citenamefont {Nguyen}, \citenamefont {Alderete}, \citenamefont
  {Perdomo-Ortiz}, \citenamefont {Korda}, \citenamefont {Garfoot},
  \citenamefont {Brecque} \emph {et~al.}}]{zhu2019training}%
  \BibitemOpen
  \bibfield  {author} {\bibinfo {author} {\bibfnamefont {D.}~\bibnamefont
  {Zhu}}, \bibinfo {author} {\bibfnamefont {N.~M.}\ \bibnamefont {Linke}},
  \bibinfo {author} {\bibfnamefont {M.}~\bibnamefont {Benedetti}}, \bibinfo
  {author} {\bibfnamefont {K.~A.}\ \bibnamefont {Landsman}}, \bibinfo {author}
  {\bibfnamefont {N.~H.}\ \bibnamefont {Nguyen}}, \bibinfo {author}
  {\bibfnamefont {C.~H.}\ \bibnamefont {Alderete}}, \bibinfo {author}
  {\bibfnamefont {A.}~\bibnamefont {Perdomo-Ortiz}}, \bibinfo {author}
  {\bibfnamefont {N.}~\bibnamefont {Korda}}, \bibinfo {author} {\bibfnamefont
  {A.}~\bibnamefont {Garfoot}}, \bibinfo {author} {\bibfnamefont
  {C.}~\bibnamefont {Brecque}}, \emph {et~al.},\ }\bibfield  {title} {\bibinfo
  {title} {Training of quantum circuits on a hybrid quantum computer},\
  }\href@noop {} {\bibfield  {journal} {\bibinfo  {journal} {Sci. Adv.}\
  }\textbf {\bibinfo {volume} {5}},\ \bibinfo {pages} {eaaw9918} (\bibinfo
  {year} {2019})}\BibitemShut {NoStop}%
\bibitem [{\citenamefont {Gheorghiu}\ \emph {et~al.}(2019)\citenamefont
  {Gheorghiu}, \citenamefont {Kapourniotis},\ and\ \citenamefont
  {Kashefi}}]{gheorghiu2019verification}%
  \BibitemOpen
  \bibfield  {author} {\bibinfo {author} {\bibfnamefont {A.}~\bibnamefont
  {Gheorghiu}}, \bibinfo {author} {\bibfnamefont {T.}~\bibnamefont
  {Kapourniotis}},\ and\ \bibinfo {author} {\bibfnamefont {E.}~\bibnamefont
  {Kashefi}},\ }\bibfield  {title} {\bibinfo {title} {Verification of quantum
  computation: An overview of existing approaches},\ }\href@noop {} {\bibfield
  {journal} {\bibinfo  {journal} {Theory Comput. Syst.}\ }\textbf {\bibinfo
  {volume} {63}},\ \bibinfo {pages} {715} (\bibinfo {year} {2019})}\BibitemShut
  {NoStop}%
\bibitem [{\citenamefont {Eisert}\ \emph {et~al.}(2020)\citenamefont {Eisert},
  \citenamefont {Hangleiter}, \citenamefont {Walk}, \citenamefont {Roth},
  \citenamefont {Markham}, \citenamefont {Parekh}, \citenamefont {Chabaud},\
  and\ \citenamefont {Kashefi}}]{eisert2020quantum}%
  \BibitemOpen
  \bibfield  {author} {\bibinfo {author} {\bibfnamefont {J.}~\bibnamefont
  {Eisert}}, \bibinfo {author} {\bibfnamefont {D.}~\bibnamefont {Hangleiter}},
  \bibinfo {author} {\bibfnamefont {N.}~\bibnamefont {Walk}}, \bibinfo {author}
  {\bibfnamefont {I.}~\bibnamefont {Roth}}, \bibinfo {author} {\bibfnamefont
  {D.}~\bibnamefont {Markham}}, \bibinfo {author} {\bibfnamefont
  {R.}~\bibnamefont {Parekh}}, \bibinfo {author} {\bibfnamefont
  {U.}~\bibnamefont {Chabaud}},\ and\ \bibinfo {author} {\bibfnamefont
  {E.}~\bibnamefont {Kashefi}},\ }\bibfield  {title} {\bibinfo {title} {Quantum
  certification and benchmarking},\ }\href@noop {} {\bibfield  {journal}
  {\bibinfo  {journal} {Nat. Rev. Phys.}\ }\textbf {\bibinfo {volume} {2}},\
  \bibinfo {pages} {382} (\bibinfo {year} {2020})}\BibitemShut {NoStop}%
\bibitem [{\citenamefont {Semnanian}\ \emph {et~al.}(2011)\citenamefont
  {Semnanian}, \citenamefont {Pham}, \citenamefont {Englert},\ and\
  \citenamefont {Wu}}]{semnanian2011virtualization}%
  \BibitemOpen
  \bibfield  {author} {\bibinfo {author} {\bibfnamefont {A.~A.}\ \bibnamefont
  {Semnanian}}, \bibinfo {author} {\bibfnamefont {J.}~\bibnamefont {Pham}},
  \bibinfo {author} {\bibfnamefont {B.}~\bibnamefont {Englert}},\ and\ \bibinfo
  {author} {\bibfnamefont {X.}~\bibnamefont {Wu}},\ }\bibfield  {title}
  {\bibinfo {title} {Virtualization technology and its impact on computer
  hardware architecture},\ }in\ \href@noop {} {\emph {\bibinfo {booktitle} {8th
  Int. Conf. Inf. Technol. Proc.}}}\ (\bibinfo {organization} {IEEE},\ \bibinfo
  {year} {2011})\ pp.\ \bibinfo {pages} {719--724}\BibitemShut {NoStop}%
\bibitem [{\citenamefont {Bechtold}(2021)}]{bechtold2021bringing}%
  \BibitemOpen
  \bibfield  {author} {\bibinfo {author} {\bibfnamefont {M.}~\bibnamefont
  {Bechtold}},\ }\href@noop {} {Master's thesis} (\bibinfo {year} {2021}),\
  \bibinfo {note} {{B}ringing the concepts of virtualization to gate-based
  quantum computing}\BibitemShut {NoStop}%
\bibitem [{QMW(2022)}]{QMWare}%
  \BibitemOpen
  \href@noop {} {\bibinfo {title} {{QM}ware: {T}he first global quantum cloud,
  https://qm-ware.com}} (\bibinfo {year} {2022})\BibitemShut {NoStop}%
\bibitem [{\citenamefont {Guerreschi}\ \emph {et~al.}(2020)\citenamefont
  {Guerreschi}, \citenamefont {Hogaboam}, \citenamefont {Baruffa},\ and\
  \citenamefont {Sawaya}}]{Guerreschi2020}%
  \BibitemOpen
  \bibfield  {author} {\bibinfo {author} {\bibfnamefont {G.~G.}\ \bibnamefont
  {Guerreschi}}, \bibinfo {author} {\bibfnamefont {J.}~\bibnamefont
  {Hogaboam}}, \bibinfo {author} {\bibfnamefont {F.}~\bibnamefont {Baruffa}},\
  and\ \bibinfo {author} {\bibfnamefont {N.~P.~D.}\ \bibnamefont {Sawaya}},\
  }\bibfield  {title} {\bibinfo {title} {Intel quantum simulator: a cloud-ready
  high-performance simulator of quantum circuits},\ }\href@noop {} {\bibfield
  {journal} {\bibinfo  {journal} {Quantum Sci. Technol.}\ }\textbf {\bibinfo
  {volume} {5}},\ \bibinfo {pages} {034007} (\bibinfo {year}
  {2020})}\BibitemShut {NoStop}%
\bibitem [{\citenamefont {Mandr{\`a}}\ \emph {et~al.}(2021)\citenamefont
  {Mandr{\`a}}, \citenamefont {Marshall}, \citenamefont {Rieffel},\ and\
  \citenamefont {Biswas}}]{mandra2021hybridq}%
  \BibitemOpen
  \bibfield  {author} {\bibinfo {author} {\bibfnamefont {S.}~\bibnamefont
  {Mandr{\`a}}}, \bibinfo {author} {\bibfnamefont {J.}~\bibnamefont
  {Marshall}}, \bibinfo {author} {\bibfnamefont {E.~G.}\ \bibnamefont
  {Rieffel}},\ and\ \bibinfo {author} {\bibfnamefont {R.}~\bibnamefont
  {Biswas}},\ }\bibfield  {title} {\bibinfo {title} {Hybridq: A hybrid
  simulator for quantum circuits},\ }in\ \href@noop {} {\emph {\bibinfo
  {booktitle} {2nd Int. Workshop on Quantum Comput. Softw.}}}\ (\bibinfo
  {organization} {IEEE/ACM},\ \bibinfo {year} {2021})\ pp.\ \bibinfo {pages}
  {99--109}\BibitemShut {NoStop}%
\bibitem [{ATO(2022)}]{ATOS}%
  \BibitemOpen
  \href@noop {} {\bibinfo {title} {{ATOS} {Q}uantum {L}earning {M}achine,
  https://atos.net/en/solutions/quantum-learning-machine}} (\bibinfo {year}
  {2022})\BibitemShut {NoStop}%
\bibitem [{\citenamefont {Efthymiou}\ \emph {et~al.}(2021)\citenamefont
  {Efthymiou}, \citenamefont {Ramos-Calderer}, \citenamefont {Bravo-Prieto},
  \citenamefont {P{\'e}rez-Salinas}, \citenamefont {Garc{\'\i}a-Mart{\'\i}n},
  \citenamefont {Garcia-Saez}, \citenamefont {Latorre},\ and\ \citenamefont
  {Carrazza}}]{efthymiou2021qibo}%
  \BibitemOpen
  \bibfield  {author} {\bibinfo {author} {\bibfnamefont {S.}~\bibnamefont
  {Efthymiou}}, \bibinfo {author} {\bibfnamefont {S.}~\bibnamefont
  {Ramos-Calderer}}, \bibinfo {author} {\bibfnamefont {C.}~\bibnamefont
  {Bravo-Prieto}}, \bibinfo {author} {\bibfnamefont {A.}~\bibnamefont
  {P{\'e}rez-Salinas}}, \bibinfo {author} {\bibfnamefont {D.}~\bibnamefont
  {Garc{\'\i}a-Mart{\'\i}n}}, \bibinfo {author} {\bibfnamefont
  {A.}~\bibnamefont {Garcia-Saez}}, \bibinfo {author} {\bibfnamefont {J.~I.}\
  \bibnamefont {Latorre}},\ and\ \bibinfo {author} {\bibfnamefont
  {S.}~\bibnamefont {Carrazza}},\ }\bibfield  {title} {\bibinfo {title} {Qibo:
  a framework for quantum simulation with hardware acceleration},\ }\href@noop
  {} {\bibfield  {journal} {\bibinfo  {journal} {Quantum Sci. Technol.}\
  }\textbf {\bibinfo {volume} {7}},\ \bibinfo {pages} {015018} (\bibinfo {year}
  {2021})}\BibitemShut {NoStop}%
\bibitem [{\citenamefont {Von~Neumann}(1993)}]{von1993first}%
  \BibitemOpen
  \bibfield  {author} {\bibinfo {author} {\bibfnamefont {J.}~\bibnamefont
  {Von~Neumann}},\ }\bibfield  {title} {\bibinfo {title} {First draft of a
  report on the {EDVAC}},\ }\href@noop {} {\bibfield  {journal} {\bibinfo
  {journal} {IEEE Ann. Hist. Comput.}\ }\textbf {\bibinfo {volume} {15}},\
  \bibinfo {pages} {27} (\bibinfo {year} {1993})}\BibitemShut {NoStop}%
\bibitem [{\citenamefont {Mariantoni}\ \emph {et~al.}(2011)\citenamefont
  {Mariantoni}, \citenamefont {Wang}, \citenamefont {Yamamoto}, \citenamefont
  {Neeley}, \citenamefont {Bialczak}, \citenamefont {Chen}, \citenamefont
  {Lenander}, \citenamefont {Lucero}, \citenamefont {O’Connell},
  \citenamefont {Sank} \emph {et~al.}}]{mariantoni2011implementing}%
  \BibitemOpen
  \bibfield  {author} {\bibinfo {author} {\bibfnamefont {M.}~\bibnamefont
  {Mariantoni}}, \bibinfo {author} {\bibfnamefont {H.}~\bibnamefont {Wang}},
  \bibinfo {author} {\bibfnamefont {T.}~\bibnamefont {Yamamoto}}, \bibinfo
  {author} {\bibfnamefont {M.}~\bibnamefont {Neeley}}, \bibinfo {author}
  {\bibfnamefont {R.~C.}\ \bibnamefont {Bialczak}}, \bibinfo {author}
  {\bibfnamefont {Y.}~\bibnamefont {Chen}}, \bibinfo {author} {\bibfnamefont
  {M.}~\bibnamefont {Lenander}}, \bibinfo {author} {\bibfnamefont
  {E.}~\bibnamefont {Lucero}}, \bibinfo {author} {\bibfnamefont {A.~D.}\
  \bibnamefont {O’Connell}}, \bibinfo {author} {\bibfnamefont
  {D.}~\bibnamefont {Sank}}, \emph {et~al.},\ }\bibfield  {title} {\bibinfo
  {title} {Implementing the quantum von neumann architecture with
  superconducting circuits},\ }\href@noop {} {\bibfield  {journal} {\bibinfo
  {journal} {Science}\ }\textbf {\bibinfo {volume} {334}},\ \bibinfo {pages}
  {61} (\bibinfo {year} {2011})}\BibitemShut {NoStop}%
\bibitem [{QIR(2022)}]{QIR}%
  \BibitemOpen
  \href@noop {} {\bibinfo {title} {{QIR} {A}lliance,
  https://github.com/qir-alliance}} (\bibinfo {year} {2022})\BibitemShut
  {NoStop}%
\bibitem [{\citenamefont {Shor}(1994)}]{Shor}%
  \BibitemOpen
  \bibfield  {author} {\bibinfo {author} {\bibfnamefont {P.~W.}\ \bibnamefont
  {Shor}},\ }\bibfield  {title} {\bibinfo {title} {Algorithms for quantum
  computation: discrete logarithms and factoring},\ }in\ \href@noop {} {\emph
  {\bibinfo {booktitle} {Proceedings 35th ASFCS}}}\ (\bibinfo  {publisher}
  {{IEEE} Comput. Soc. Press},\ \bibinfo {year} {1994})\BibitemShut {NoStop}%
\bibitem [{\citenamefont {Cerezo}\ \emph {et~al.}(2021)\citenamefont {Cerezo},
  \citenamefont {Arrasmith}, \citenamefont {Babbush}, \citenamefont {Benjamin},
  \citenamefont {Endo}, \citenamefont {Fujii}, \citenamefont {McClean},
  \citenamefont {Mitarai}, \citenamefont {Yuan}, \citenamefont {Cincio},\ and\
  \citenamefont {Coles}}]{VQC2021}%
  \BibitemOpen
  \bibfield  {author} {\bibinfo {author} {\bibfnamefont {M.}~\bibnamefont
  {Cerezo}}, \bibinfo {author} {\bibfnamefont {A.}~\bibnamefont {Arrasmith}},
  \bibinfo {author} {\bibfnamefont {R.}~\bibnamefont {Babbush}}, \bibinfo
  {author} {\bibfnamefont {S.~C.}\ \bibnamefont {Benjamin}}, \bibinfo {author}
  {\bibfnamefont {S.}~\bibnamefont {Endo}}, \bibinfo {author} {\bibfnamefont
  {K.}~\bibnamefont {Fujii}}, \bibinfo {author} {\bibfnamefont {J.~R.}\
  \bibnamefont {McClean}}, \bibinfo {author} {\bibfnamefont {K.}~\bibnamefont
  {Mitarai}}, \bibinfo {author} {\bibfnamefont {X.}~\bibnamefont {Yuan}},
  \bibinfo {author} {\bibfnamefont {L.}~\bibnamefont {Cincio}},\ and\ \bibinfo
  {author} {\bibfnamefont {P.~J.}\ \bibnamefont {Coles}},\ }\bibfield  {title}
  {\bibinfo {title} {Variational quantum algorithms},\ }\href@noop {}
  {\bibfield  {journal} {\bibinfo  {journal} {Nat. Rev. Phys.}\ }\textbf
  {\bibinfo {volume} {3}},\ \bibinfo {pages} {625} (\bibinfo {year}
  {2021})}\BibitemShut {NoStop}%
\bibitem [{\citenamefont {Shi}\ \emph {et~al.}(2006)\citenamefont {Shi},
  \citenamefont {Duan},\ and\ \citenamefont {Vidal}}]{TN2006}%
  \BibitemOpen
  \bibfield  {author} {\bibinfo {author} {\bibfnamefont {Y.-Y.}\ \bibnamefont
  {Shi}}, \bibinfo {author} {\bibfnamefont {L.-M.}\ \bibnamefont {Duan}},\ and\
  \bibinfo {author} {\bibfnamefont {G.}~\bibnamefont {Vidal}},\ }\bibfield
  {title} {\bibinfo {title} {Classical simulation of quantum many-body systems
  with a tree tensor network},\ }\href@noop {} {\bibfield  {journal} {\bibinfo
  {journal} {Phys. Rev. A}\ }\textbf {\bibinfo {volume} {74}} (\bibinfo {year}
  {2006})}\BibitemShut {NoStop}%
\bibitem [{\citenamefont {Niwa}\ \emph {et~al.}(2002)\citenamefont {Niwa},
  \citenamefont {Matsumoto},\ and\ \citenamefont {Imai}}]{Niwa2002}%
  \BibitemOpen
  \bibfield  {author} {\bibinfo {author} {\bibfnamefont {J.}~\bibnamefont
  {Niwa}}, \bibinfo {author} {\bibfnamefont {K.}~\bibnamefont {Matsumoto}},\
  and\ \bibinfo {author} {\bibfnamefont {H.}~\bibnamefont {Imai}},\ }\bibfield
  {title} {\bibinfo {title} {General-purpose parallel simulator for quantum
  computing},\ }in\ \href@noop {} {\emph {\bibinfo {booktitle} {Unconventional
  Models of Computation}}}\ (\bibinfo  {publisher} {Springer Berlin
  Heidelberg},\ \bibinfo {year} {2002})\ pp.\ \bibinfo {pages}
  {230--251}\BibitemShut {NoStop}%
\bibitem [{\citenamefont {Jones}\ \emph {et~al.}(2019)\citenamefont {Jones},
  \citenamefont {Brown}, \citenamefont {Bush},\ and\ \citenamefont
  {Benjamin}}]{Jones2019}%
  \BibitemOpen
  \bibfield  {author} {\bibinfo {author} {\bibfnamefont {T.}~\bibnamefont
  {Jones}}, \bibinfo {author} {\bibfnamefont {A.}~\bibnamefont {Brown}},
  \bibinfo {author} {\bibfnamefont {I.}~\bibnamefont {Bush}},\ and\ \bibinfo
  {author} {\bibfnamefont {S.~C.}\ \bibnamefont {Benjamin}},\ }\bibfield
  {title} {\bibinfo {title} {{QuEST} and high performance simulation of quantum
  computers},\ }\href@noop {} {\bibfield  {journal} {\bibinfo  {journal} {Sci.
  Rep.}\ }\textbf {\bibinfo {volume} {9}} (\bibinfo {year} {2019})}\BibitemShut
  {NoStop}%
\bibitem [{\citenamefont {Venegas-Andraca}\ \emph {et~al.}(2018)\citenamefont
  {Venegas-Andraca}, \citenamefont {Cruz-Santos}, \citenamefont {McGeoch},\
  and\ \citenamefont {Lanzagorta}}]{Salvador2018}%
  \BibitemOpen
  \bibfield  {author} {\bibinfo {author} {\bibfnamefont {S.~E.}\ \bibnamefont
  {Venegas-Andraca}}, \bibinfo {author} {\bibfnamefont {W.}~\bibnamefont
  {Cruz-Santos}}, \bibinfo {author} {\bibfnamefont {C.}~\bibnamefont
  {McGeoch}},\ and\ \bibinfo {author} {\bibfnamefont {M.}~\bibnamefont
  {Lanzagorta}},\ }\bibfield  {title} {\bibinfo {title} {A cross-disciplinary
  introduction to quantum annealing-based algorithms},\ }\href@noop {}
  {\bibfield  {journal} {\bibinfo  {journal} {Contemp. Phys.}\ }\textbf
  {\bibinfo {volume} {59}},\ \bibinfo {pages} {174} (\bibinfo {year}
  {2018})}\BibitemShut {NoStop}%
\bibitem [{\citenamefont {Farhi}\ \emph {et~al.}(2014)\citenamefont {Farhi},
  \citenamefont {Goldstone},\ and\ \citenamefont {Gutmann}}]{QAOA}%
  \BibitemOpen
  \bibfield  {author} {\bibinfo {author} {\bibfnamefont {E.}~\bibnamefont
  {Farhi}}, \bibinfo {author} {\bibfnamefont {J.}~\bibnamefont {Goldstone}},\
  and\ \bibinfo {author} {\bibfnamefont {S.}~\bibnamefont {Gutmann}},\
  }\href@noop {} {\bibinfo {title} {A quantum approximate optimization
  algorithm}},\ \bibinfo {howpublished} {arXiv:1411.4028} (\bibinfo {year}
  {2014})\BibitemShut {NoStop}%
\bibitem [{\citenamefont {Pillonetto}\ \emph {et~al.}(2014)\citenamefont
  {Pillonetto}, \citenamefont {Dinuzzo}, \citenamefont {Chen}, \citenamefont
  {Nicolao},\ and\ \citenamefont {Ljung}}]{Kernel2014}%
  \BibitemOpen
  \bibfield  {author} {\bibinfo {author} {\bibfnamefont {G.}~\bibnamefont
  {Pillonetto}}, \bibinfo {author} {\bibfnamefont {F.}~\bibnamefont {Dinuzzo}},
  \bibinfo {author} {\bibfnamefont {T.}~\bibnamefont {Chen}}, \bibinfo {author}
  {\bibfnamefont {G.~D.}\ \bibnamefont {Nicolao}},\ and\ \bibinfo {author}
  {\bibfnamefont {L.}~\bibnamefont {Ljung}},\ }\bibfield  {title} {\bibinfo
  {title} {Kernel methods in system identification, machine learning and
  function estimation: A survey},\ }\href@noop {} {\bibfield  {journal}
  {\bibinfo  {journal} {Automatica}\ }\textbf {\bibinfo {volume} {50}},\
  \bibinfo {pages} {657} (\bibinfo {year} {2014})}\BibitemShut {NoStop}%
\bibitem [{\citenamefont {Liu}\ \emph {et~al.}(2021)\citenamefont {Liu},
  \citenamefont {Arunachalam},\ and\ \citenamefont {Temme}}]{ML2021}%
  \BibitemOpen
  \bibfield  {author} {\bibinfo {author} {\bibfnamefont {Y.}~\bibnamefont
  {Liu}}, \bibinfo {author} {\bibfnamefont {S.}~\bibnamefont {Arunachalam}},\
  and\ \bibinfo {author} {\bibfnamefont {K.}~\bibnamefont {Temme}},\ }\bibfield
   {title} {\bibinfo {title} {A rigorous and robust quantum speed-up in
  supervised machine learning},\ }\href@noop {} {\bibfield  {journal} {\bibinfo
   {journal} {Nat. Phys.}\ }\textbf {\bibinfo {volume} {17}},\ \bibinfo {pages}
  {1013} (\bibinfo {year} {2021})}\BibitemShut {NoStop}%
\bibitem [{\citenamefont {Arute}\ \emph {et~al.}(2019)\citenamefont {Arute},
  \citenamefont {Arya}, \citenamefont {Babbush}, \citenamefont {Bacon},
  \citenamefont {Bardin} \emph {et~al.}}]{Arute2019}%
  \BibitemOpen
  \bibfield  {author} {\bibinfo {author} {\bibfnamefont {F.}~\bibnamefont
  {Arute}}, \bibinfo {author} {\bibfnamefont {K.}~\bibnamefont {Arya}},
  \bibinfo {author} {\bibfnamefont {R.}~\bibnamefont {Babbush}}, \bibinfo
  {author} {\bibfnamefont {D.}~\bibnamefont {Bacon}}, \bibinfo {author}
  {\bibfnamefont {J.~C.}\ \bibnamefont {Bardin}}, \emph {et~al.},\ }\bibfield
  {title} {\bibinfo {title} {Quantum supremacy using a programmable
  superconducting processor},\ }\href@noop {} {\bibfield  {journal} {\bibinfo
  {journal} {Nature}\ }\textbf {\bibinfo {volume} {574}},\ \bibinfo {pages}
  {505} (\bibinfo {year} {2019})}\BibitemShut {NoStop}%
\bibitem [{\citenamefont {Wu}\ \emph {et~al.}(2021)\citenamefont {Wu},
  \citenamefont {Bao}, \citenamefont {Cao}, \citenamefont {Chen}, \citenamefont
  {Chen},\ and\ \citenamefont {Chen}}]{SupremacyChina2021a}%
  \BibitemOpen
  \bibfield  {author} {\bibinfo {author} {\bibfnamefont {Y.}~\bibnamefont
  {Wu}}, \bibinfo {author} {\bibfnamefont {W.-S.}\ \bibnamefont {Bao}},
  \bibinfo {author} {\bibfnamefont {S.}~\bibnamefont {Cao}}, \bibinfo {author}
  {\bibfnamefont {F.}~\bibnamefont {Chen}}, \bibinfo {author} {\bibfnamefont
  {M.-C.}\ \bibnamefont {Chen}},\ and\ \bibinfo {author} {\bibfnamefont
  {X.}~\bibnamefont {Chen}},\ }\bibfield  {title} {\bibinfo {title} {Strong
  quantum computational advantage using a superconducting quantum processor},\
  }\href@noop {} {\bibfield  {journal} {\bibinfo  {journal} {arXiv:2106.14734}\
  } (\bibinfo {year} {2021})}\BibitemShut {NoStop}%
\bibitem [{\citenamefont {Zhu}\ \emph {et~al.}(2021)\citenamefont {Zhu},
  \citenamefont {Cao}, \citenamefont {Chen}, \citenamefont {Chen.},
  \citenamefont {Chen},\ and\ \citenamefont {Chung}}]{SupremacyChina2021b}%
  \BibitemOpen
  \bibfield  {author} {\bibinfo {author} {\bibfnamefont {Q.}~\bibnamefont
  {Zhu}}, \bibinfo {author} {\bibfnamefont {S.}~\bibnamefont {Cao}}, \bibinfo
  {author} {\bibfnamefont {F.}~\bibnamefont {Chen}}, \bibinfo {author}
  {\bibfnamefont {M.-C.}\ \bibnamefont {Chen.}}, \bibinfo {author}
  {\bibfnamefont {X.}~\bibnamefont {Chen}},\ and\ \bibinfo {author}
  {\bibfnamefont {T.-H.}\ \bibnamefont {Chung}},\ }\bibfield  {title} {\bibinfo
  {title} {Quantum computational advantage via 60-qubit 24-cycle random circuit
  sampling},\ }\href@noop {} {\bibfield  {journal} {\bibinfo  {journal}
  {arXiv:2109.03494}\ } (\bibinfo {year} {2021})}\BibitemShut {NoStop}%
\bibitem [{\citenamefont {Perelshtein}\ \emph {et~al.}(2020)\citenamefont
  {Perelshtein}, \citenamefont {Pakhomchik}, \citenamefont {Melnikov},
  \citenamefont {Novikov}, \citenamefont {Glatz}, \citenamefont {Paraoanu},
  \citenamefont {Vinokur},\ and\ \citenamefont {Lesovik}}]{Perelshtein2020}%
  \BibitemOpen
  \bibfield  {author} {\bibinfo {author} {\bibfnamefont {M.~R.}\ \bibnamefont
  {Perelshtein}}, \bibinfo {author} {\bibfnamefont {A.~I.}\ \bibnamefont
  {Pakhomchik}}, \bibinfo {author} {\bibfnamefont {A.~A.}\ \bibnamefont
  {Melnikov}}, \bibinfo {author} {\bibfnamefont {A.~A.}\ \bibnamefont
  {Novikov}}, \bibinfo {author} {\bibfnamefont {A.}~\bibnamefont {Glatz}},
  \bibinfo {author} {\bibfnamefont {G.~S.}\ \bibnamefont {Paraoanu}}, \bibinfo
  {author} {\bibfnamefont {V.~M.}\ \bibnamefont {Vinokur}},\ and\ \bibinfo
  {author} {\bibfnamefont {G.~B.}\ \bibnamefont {Lesovik}},\ }\bibfield
  {title} {\bibinfo {title} {Solving large-scale linear systems of equations by
  a quantum hybrid algorithm},\ }\href@noop {} {\bibfield  {journal} {\bibinfo
  {journal} {arXiv:2003.12770}\ } (\bibinfo {year} {2020})}\BibitemShut
  {NoStop}%
\bibitem [{\citenamefont {Perelshtein}\ \emph {et~al.}(2021)\citenamefont
  {Perelshtein}, \citenamefont {Kirsanov}, \citenamefont {Zemlyanov},
  \citenamefont {Lebedev}, \citenamefont {Blatter}, \citenamefont {Vinokur},\
  and\ \citenamefont {Lesovik}}]{LAMA}%
  \BibitemOpen
  \bibfield  {author} {\bibinfo {author} {\bibfnamefont {M.~R.}\ \bibnamefont
  {Perelshtein}}, \bibinfo {author} {\bibfnamefont {N.~S.}\ \bibnamefont
  {Kirsanov}}, \bibinfo {author} {\bibfnamefont {V.~V.}\ \bibnamefont
  {Zemlyanov}}, \bibinfo {author} {\bibfnamefont {A.~V.}\ \bibnamefont
  {Lebedev}}, \bibinfo {author} {\bibfnamefont {G.}~\bibnamefont {Blatter}},
  \bibinfo {author} {\bibfnamefont {V.~M.}\ \bibnamefont {Vinokur}},\ and\
  \bibinfo {author} {\bibfnamefont {G.~B.}\ \bibnamefont {Lesovik}},\
  }\bibfield  {title} {\bibinfo {title} {Linear ascending metrological
  algorithm},\ }\href@noop {} {\bibfield  {journal} {\bibinfo  {journal} {Phys.
  Rev. Research}\ }\textbf {\bibinfo {volume} {3}},\ \bibinfo {pages} {013257}
  (\bibinfo {year} {2021})}\BibitemShut {NoStop}%
\bibitem [{\citenamefont {Cheeseman}\ \emph {et~al.}(1991)\citenamefont
  {Cheeseman}, \citenamefont {Kanefsky}, \citenamefont {Taylor} \emph
  {et~al.}}]{cheeseman1991really}%
  \BibitemOpen
  \bibfield  {author} {\bibinfo {author} {\bibfnamefont {P.~C.}\ \bibnamefont
  {Cheeseman}}, \bibinfo {author} {\bibfnamefont {B.}~\bibnamefont {Kanefsky}},
  \bibinfo {author} {\bibfnamefont {W.~M.}\ \bibnamefont {Taylor}}, \emph
  {et~al.},\ }\bibfield  {title} {\bibinfo {title} {Where the really hard
  problems are},\ }in\ \href@noop {} {\emph {\bibinfo {booktitle} {IJCAI}}},\
  Vol.~\bibinfo {volume} {91}\ (\bibinfo {year} {1991})\ pp.\ \bibinfo {pages}
  {331--337}\BibitemShut {NoStop}%
\bibitem [{\citenamefont {Strubell}\ \emph {et~al.}(2019)\citenamefont
  {Strubell}, \citenamefont {Ganesh},\ and\ \citenamefont
  {McCallum}}]{strubell2019energy}%
  \BibitemOpen
  \bibfield  {author} {\bibinfo {author} {\bibfnamefont {E.}~\bibnamefont
  {Strubell}}, \bibinfo {author} {\bibfnamefont {A.}~\bibnamefont {Ganesh}},\
  and\ \bibinfo {author} {\bibfnamefont {A.}~\bibnamefont {McCallum}},\
  }\bibfield  {title} {\bibinfo {title} {Energy and policy considerations for
  deep learning in {NLP}},\ }\href@noop {} {\bibfield  {journal} {\bibinfo
  {journal} {arXiv:1906.02243}\ } (\bibinfo {year} {2019})}\BibitemShut
  {NoStop}%
\bibitem [{\citenamefont {Georgescu}\ \emph {et~al.}(2014)\citenamefont
  {Georgescu}, \citenamefont {Ashhab},\ and\ \citenamefont
  {Nori}}]{georgescu2014quantum}%
  \BibitemOpen
  \bibfield  {author} {\bibinfo {author} {\bibfnamefont {I.~M.}\ \bibnamefont
  {Georgescu}}, \bibinfo {author} {\bibfnamefont {S.}~\bibnamefont {Ashhab}},\
  and\ \bibinfo {author} {\bibfnamefont {F.}~\bibnamefont {Nori}},\ }\bibfield
  {title} {\bibinfo {title} {Quantum simulation},\ }\href@noop {} {\bibfield
  {journal} {\bibinfo  {journal} {Reviews of Modern Physics}\ }\textbf
  {\bibinfo {volume} {86}},\ \bibinfo {pages} {153} (\bibinfo {year}
  {2014})}\BibitemShut {NoStop}%
\bibitem [{\citenamefont {Karp}(1972)}]{Karp1972}%
  \BibitemOpen
  \bibfield  {author} {\bibinfo {author} {\bibfnamefont {R.~M.}\ \bibnamefont
  {Karp}},\ }\bibfield  {title} {\bibinfo {title} {Reducibility among
  combinatorial problems}\ }(\bibinfo  {publisher} {Springer {US}},\ \bibinfo
  {year} {1972})\ pp.\ \bibinfo {pages} {85--103}\BibitemShut {NoStop}%
\bibitem [{\citenamefont {Lawler}\ and\ \citenamefont {Wood}(1966)}]{1966}%
  \BibitemOpen
  \bibfield  {author} {\bibinfo {author} {\bibfnamefont {E.~L.}\ \bibnamefont
  {Lawler}}\ and\ \bibinfo {author} {\bibfnamefont {D.~E.}\ \bibnamefont
  {Wood}},\ }\bibfield  {title} {\bibinfo {title} {Branch-and-{B}ound methods:
  A survey},\ }\href@noop {} {\bibfield  {journal} {\bibinfo  {journal} {Oper.
  Res.}\ }\textbf {\bibinfo {volume} {14}},\ \bibinfo {pages} {699} (\bibinfo
  {year} {1966})}\BibitemShut {NoStop}%
\bibitem [{\citenamefont {Dongarra}\ and\ \citenamefont
  {Sullivan}(2000)}]{Dongarra2000}%
  \BibitemOpen
  \bibfield  {author} {\bibinfo {author} {\bibfnamefont {J.}~\bibnamefont
  {Dongarra}}\ and\ \bibinfo {author} {\bibfnamefont {F.}~\bibnamefont
  {Sullivan}},\ }\bibfield  {title} {\bibinfo {title} {Guest editors'
  introduction to the top 10 algorithms},\ }\href@noop {} {\bibfield  {journal}
  {\bibinfo  {journal} {IEEE Comput. Archit. Lett.}\ }\textbf {\bibinfo
  {volume} {2}},\ \bibinfo {pages} {22} (\bibinfo {year} {2000})}\BibitemShut
  {NoStop}%
\bibitem [{\citenamefont {{IBM ILOG CPLEX}}(2009)}]{cplex2009v12}%
  \BibitemOpen
  \bibfield  {author} {\bibinfo {author} {\bibnamefont {{IBM ILOG CPLEX}}},\
  }\bibfield  {title} {\bibinfo {title} {User’s manual for {CPLEX}},\
  }\href@noop {} {\bibfield  {journal} {\bibinfo  {journal} {International
  Business Machines Corporation}\ }\textbf {\bibinfo {volume} {46}},\ \bibinfo
  {pages} {157} (\bibinfo {year} {2009})}\BibitemShut {NoStop}%
\bibitem [{\citenamefont {{Gurobi Optimization, LLC}}(2021)}]{gurobi}%
  \BibitemOpen
  \bibfield  {author} {\bibinfo {author} {\bibnamefont {{Gurobi Optimization,
  LLC}}},\ }\href@noop {} {\bibinfo {title} {{Gurobi Optimizer Reference
  Manual}}} (\bibinfo {year} {2021})\BibitemShut {NoStop}%
\bibitem [{\citenamefont {Knuth}(1974)}]{Knuth1974}%
  \BibitemOpen
  \bibfield  {author} {\bibinfo {author} {\bibfnamefont {D.~E.}\ \bibnamefont
  {Knuth}},\ }\bibfield  {title} {\bibinfo {title} {Postscript about {NP}-hard
  problems},\ }\href@noop {} {\bibfield  {journal} {\bibinfo  {journal} {ACM
  SIGACT News}\ }\textbf {\bibinfo {volume} {6}},\ \bibinfo {pages} {15}
  (\bibinfo {year} {1974})}\BibitemShut {NoStop}%
\bibitem [{\citenamefont {Farhi}\ \emph {et~al.}(2001)\citenamefont {Farhi},
  \citenamefont {Goldstone}, \citenamefont {Gutmann}, \citenamefont {Lapan},
  \citenamefont {Lundgren},\ and\ \citenamefont {Preda}}]{farhi2001quantum}%
  \BibitemOpen
  \bibfield  {author} {\bibinfo {author} {\bibfnamefont {E.}~\bibnamefont
  {Farhi}}, \bibinfo {author} {\bibfnamefont {J.}~\bibnamefont {Goldstone}},
  \bibinfo {author} {\bibfnamefont {S.}~\bibnamefont {Gutmann}}, \bibinfo
  {author} {\bibfnamefont {J.}~\bibnamefont {Lapan}}, \bibinfo {author}
  {\bibfnamefont {A.}~\bibnamefont {Lundgren}},\ and\ \bibinfo {author}
  {\bibfnamefont {D.}~\bibnamefont {Preda}},\ }\bibfield  {title} {\bibinfo
  {title} {A quantum adiabatic evolution algorithm applied to random instances
  of an {NP}-complete problem},\ }\href@noop {} {\bibfield  {journal} {\bibinfo
   {journal} {Science}\ }\textbf {\bibinfo {volume} {292}},\ \bibinfo {pages}
  {472} (\bibinfo {year} {2001})}\BibitemShut {NoStop}%
\bibitem [{\citenamefont {Moll}\ \emph {et~al.}(2018)\citenamefont {Moll},
  \citenamefont {Barkoutsos}, \citenamefont {Bishop}, \citenamefont {Chow},
  \citenamefont {Cross}, \citenamefont {Egger}, \citenamefont {Filipp},
  \citenamefont {Fuhrer}, \citenamefont {Gambetta}, \citenamefont {Ganzhorn}
  \emph {et~al.}}]{moll2018quantum}%
  \BibitemOpen
  \bibfield  {author} {\bibinfo {author} {\bibfnamefont {N.}~\bibnamefont
  {Moll}}, \bibinfo {author} {\bibfnamefont {P.}~\bibnamefont {Barkoutsos}},
  \bibinfo {author} {\bibfnamefont {L.~S.}\ \bibnamefont {Bishop}}, \bibinfo
  {author} {\bibfnamefont {J.~M.}\ \bibnamefont {Chow}}, \bibinfo {author}
  {\bibfnamefont {A.}~\bibnamefont {Cross}}, \bibinfo {author} {\bibfnamefont
  {D.~J.}\ \bibnamefont {Egger}}, \bibinfo {author} {\bibfnamefont
  {S.}~\bibnamefont {Filipp}}, \bibinfo {author} {\bibfnamefont
  {A.}~\bibnamefont {Fuhrer}}, \bibinfo {author} {\bibfnamefont {J.~M.}\
  \bibnamefont {Gambetta}}, \bibinfo {author} {\bibfnamefont {M.}~\bibnamefont
  {Ganzhorn}}, \emph {et~al.},\ }\bibfield  {title} {\bibinfo {title} {Quantum
  optimization using variational algorithms on near-term quantum devices},\
  }\href@noop {} {\bibfield  {journal} {\bibinfo  {journal} {Quantum Sci.
  Technol.}\ }\textbf {\bibinfo {volume} {3}},\ \bibinfo {pages} {030503}
  (\bibinfo {year} {2018})}\BibitemShut {NoStop}%
\bibitem [{\citenamefont {Marsh}\ and\ \citenamefont
  {Wang}(2020)}]{PhysRevResearch.2.023302}%
  \BibitemOpen
  \bibfield  {author} {\bibinfo {author} {\bibfnamefont {S.}~\bibnamefont
  {Marsh}}\ and\ \bibinfo {author} {\bibfnamefont {J.~B.}\ \bibnamefont
  {Wang}},\ }\bibfield  {title} {\bibinfo {title} {Combinatorial optimization
  via highly efficient quantum walks},\ }\href@noop {} {\bibfield  {journal}
  {\bibinfo  {journal} {Phys. Rev. Research}\ }\textbf {\bibinfo {volume}
  {2}},\ \bibinfo {pages} {023302} (\bibinfo {year} {2020})}\BibitemShut
  {NoStop}%
\bibitem [{\citenamefont {Das}\ and\ \citenamefont {Chakrabarti}(2005)}]{2005}%
  \BibitemOpen
  \bibinfo {editor} {\bibfnamefont {A.}~\bibnamefont {Das}}\ and\ \bibinfo
  {editor} {\bibfnamefont {B.~K.}\ \bibnamefont {Chakrabarti}},\ eds.,\
  \href@noop {} {\emph {\bibinfo {title} {Quantum Annealing and Other
  Optimization Methods}}}\ (\bibinfo  {publisher} {Springer Berlin
  Heidelberg},\ \bibinfo {year} {2005})\BibitemShut {NoStop}%
\bibitem [{\citenamefont {Lucas}(2014)}]{Lucas2014}%
  \BibitemOpen
  \bibfield  {author} {\bibinfo {author} {\bibfnamefont {A.}~\bibnamefont
  {Lucas}},\ }\bibfield  {title} {\bibinfo {title} {Ising formulations of many
  {NP} problems},\ }\href@noop {} {\bibfield  {journal} {\bibinfo  {journal}
  {Front. Phys.}\ }\textbf {\bibinfo {volume} {2}} (\bibinfo {year}
  {2014})}\BibitemShut {NoStop}%
\bibitem [{\citenamefont {Ausiello}\ \emph {et~al.}(1999)\citenamefont
  {Ausiello}, \citenamefont {Marchetti-Spaccamela}, \citenamefont {Crescenzi},
  \citenamefont {Gambosi}, \citenamefont {Protasi},\ and\ \citenamefont
  {Kann}}]{Ausiello1999}%
  \BibitemOpen
  \bibfield  {author} {\bibinfo {author} {\bibfnamefont {G.}~\bibnamefont
  {Ausiello}}, \bibinfo {author} {\bibfnamefont {A.}~\bibnamefont
  {Marchetti-Spaccamela}}, \bibinfo {author} {\bibfnamefont {P.}~\bibnamefont
  {Crescenzi}}, \bibinfo {author} {\bibfnamefont {G.}~\bibnamefont {Gambosi}},
  \bibinfo {author} {\bibfnamefont {M.}~\bibnamefont {Protasi}},\ and\ \bibinfo
  {author} {\bibfnamefont {V.}~\bibnamefont {Kann}},\ }\href@noop {} {\emph
  {\bibinfo {title} {Complexity and Approximation}}}\ (\bibinfo  {publisher}
  {Springer Berlin Heidelberg},\ \bibinfo {year} {1999})\BibitemShut {NoStop}%
\bibitem [{\citenamefont {B\"{o}hm}\ \emph {et~al.}(2019)\citenamefont
  {B\"{o}hm}, \citenamefont {Verschaffelt},\ and\ \citenamefont {der
  Sande}}]{CIM2019}%
  \BibitemOpen
  \bibfield  {author} {\bibinfo {author} {\bibfnamefont {F.}~\bibnamefont
  {B\"{o}hm}}, \bibinfo {author} {\bibfnamefont {G.}~\bibnamefont
  {Verschaffelt}},\ and\ \bibinfo {author} {\bibfnamefont {G.~V.}\ \bibnamefont
  {der Sande}},\ }\bibfield  {title} {\bibinfo {title} {A poor man's coherent
  ising machine based on opto-electronic feedback systems for solving
  optimization problems},\ }\href@noop {} {\bibfield  {journal} {\bibinfo
  {journal} {Nat. Commun.}\ }\textbf {\bibinfo {volume} {10}} (\bibinfo {year}
  {2019})}\BibitemShut {NoStop}%
\bibitem [{\citenamefont {Hamerly}\ \emph {et~al.}(2019)\citenamefont
  {Hamerly}, \citenamefont {Inagaki}, \citenamefont {McMahon}, \citenamefont
  {Venturelli}, \citenamefont {Marandi} \emph {et~al.}}]{Hamerly2019}%
  \BibitemOpen
  \bibfield  {author} {\bibinfo {author} {\bibfnamefont {R.}~\bibnamefont
  {Hamerly}}, \bibinfo {author} {\bibfnamefont {T.}~\bibnamefont {Inagaki}},
  \bibinfo {author} {\bibfnamefont {P.~L.}\ \bibnamefont {McMahon}}, \bibinfo
  {author} {\bibfnamefont {D.}~\bibnamefont {Venturelli}}, \bibinfo {author}
  {\bibfnamefont {A.}~\bibnamefont {Marandi}}, \emph {et~al.},\ }\bibfield
  {title} {\bibinfo {title} {Experimental investigation of performance
  differences between coherent ising machines and a quantum annealer},\
  }\href@noop {} {\bibfield  {journal} {\bibinfo  {journal} {Sci. Adv.}\
  }\textbf {\bibinfo {volume} {5}},\ \bibinfo {pages} {eaau0823} (\bibinfo
  {year} {2019})}\BibitemShut {NoStop}%
\bibitem [{\citenamefont {Yamamoto}\ \emph {et~al.}(2017)\citenamefont
  {Yamamoto}, \citenamefont {Aihara}, \citenamefont {Leleu}, \citenamefont
  {ichi Kawarabayashi}, \citenamefont {Kako}, \citenamefont {Fejer},
  \citenamefont {Inoue},\ and\ \citenamefont {Takesue}}]{CIM2017}%
  \BibitemOpen
  \bibfield  {author} {\bibinfo {author} {\bibfnamefont {Y.}~\bibnamefont
  {Yamamoto}}, \bibinfo {author} {\bibfnamefont {K.}~\bibnamefont {Aihara}},
  \bibinfo {author} {\bibfnamefont {T.}~\bibnamefont {Leleu}}, \bibinfo
  {author} {\bibfnamefont {K.}~\bibnamefont {ichi Kawarabayashi}}, \bibinfo
  {author} {\bibfnamefont {S.}~\bibnamefont {Kako}}, \bibinfo {author}
  {\bibfnamefont {M.}~\bibnamefont {Fejer}}, \bibinfo {author} {\bibfnamefont
  {K.}~\bibnamefont {Inoue}},\ and\ \bibinfo {author} {\bibfnamefont
  {H.}~\bibnamefont {Takesue}},\ }\bibfield  {title} {\bibinfo {title}
  {Coherent ising machines {\textemdash} optical neural networks operating at
  the quantum limit},\ }\href@noop {} {\bibfield  {journal} {\bibinfo
  {journal} {npj Quantum Inf.}\ }\textbf {\bibinfo {volume} {3}} (\bibinfo
  {year} {2017})}\BibitemShut {NoStop}%
\bibitem [{\citenamefont {Harrigan}\ \emph {et~al.}(2021)\citenamefont
  {Harrigan}, \citenamefont {Sung}, \citenamefont {Neeley}, \citenamefont
  {Satzinger}, \citenamefont {Arute} \emph {et~al.}}]{Harrigan2021}%
  \BibitemOpen
  \bibfield  {author} {\bibinfo {author} {\bibfnamefont {M.~P.}\ \bibnamefont
  {Harrigan}}, \bibinfo {author} {\bibfnamefont {K.~J.}\ \bibnamefont {Sung}},
  \bibinfo {author} {\bibfnamefont {M.}~\bibnamefont {Neeley}}, \bibinfo
  {author} {\bibfnamefont {K.~J.}\ \bibnamefont {Satzinger}}, \bibinfo {author}
  {\bibfnamefont {F.}~\bibnamefont {Arute}}, \emph {et~al.},\ }\bibfield
  {title} {\bibinfo {title} {Quantum approximate optimization of non-planar
  graph problems on a planar superconducting processor},\ }\href@noop {}
  {\bibfield  {journal} {\bibinfo  {journal} {Nat. Phys.}\ }\textbf {\bibinfo
  {volume} {17}},\ \bibinfo {pages} {332} (\bibinfo {year} {2021})}\BibitemShut
  {NoStop}%
\bibitem [{\citenamefont {Amaro}\ \emph {et~al.}(2021)\citenamefont {Amaro},
  \citenamefont {Modica}, \citenamefont {Rosenkranz}, \citenamefont
  {Fiorentini}, \citenamefont {Benedetti},\ and\ \citenamefont
  {Lubasch}}]{amaro2021filtering}%
  \BibitemOpen
  \bibfield  {author} {\bibinfo {author} {\bibfnamefont {D.}~\bibnamefont
  {Amaro}}, \bibinfo {author} {\bibfnamefont {C.}~\bibnamefont {Modica}},
  \bibinfo {author} {\bibfnamefont {M.}~\bibnamefont {Rosenkranz}}, \bibinfo
  {author} {\bibfnamefont {M.}~\bibnamefont {Fiorentini}}, \bibinfo {author}
  {\bibfnamefont {M.}~\bibnamefont {Benedetti}},\ and\ \bibinfo {author}
  {\bibfnamefont {M.}~\bibnamefont {Lubasch}},\ }\bibfield  {title} {\bibinfo
  {title} {Filtering variational quantum algorithms for combinatorial
  optimization},\ }\href@noop {} {\bibfield  {journal} {\bibinfo  {journal}
  {arXiv:2106.10055}\ } (\bibinfo {year} {2021})}\BibitemShut {NoStop}%
\bibitem [{\citenamefont {Tan}\ \emph {et~al.}(2021)\citenamefont {Tan},
  \citenamefont {Lemonde}, \citenamefont {Thanasilp}, \citenamefont
  {Tangpanitanon},\ and\ \citenamefont {Angelakis}}]{Tan2021}%
  \BibitemOpen
  \bibfield  {author} {\bibinfo {author} {\bibfnamefont {B.}~\bibnamefont
  {Tan}}, \bibinfo {author} {\bibfnamefont {M.-A.}\ \bibnamefont {Lemonde}},
  \bibinfo {author} {\bibfnamefont {S.}~\bibnamefont {Thanasilp}}, \bibinfo
  {author} {\bibfnamefont {J.}~\bibnamefont {Tangpanitanon}},\ and\ \bibinfo
  {author} {\bibfnamefont {D.~G.}\ \bibnamefont {Angelakis}},\ }\bibfield
  {title} {\bibinfo {title} {Qubit-efficient encoding schemes for binary
  optimisation problems},\ }\href@noop {} {\bibfield  {journal} {\bibinfo
  {journal} {Quantum}\ }\textbf {\bibinfo {volume} {5}},\ \bibinfo {pages}
  {454} (\bibinfo {year} {2021})}\BibitemShut {NoStop}%
\bibitem [{\citenamefont {Broughton}\ \emph {et~al.}(2020)\citenamefont
  {Broughton}, \citenamefont {Verdon}, \citenamefont {McCourt}, \citenamefont
  {Martinez}, \citenamefont {Yoo} \emph {et~al.}}]{Broughton2020TensorFlowQA}%
  \BibitemOpen
  \bibfield  {author} {\bibinfo {author} {\bibfnamefont {M.}~\bibnamefont
  {Broughton}}, \bibinfo {author} {\bibfnamefont {G.}~\bibnamefont {Verdon}},
  \bibinfo {author} {\bibfnamefont {T.}~\bibnamefont {McCourt}}, \bibinfo
  {author} {\bibfnamefont {A.~J.}\ \bibnamefont {Martinez}}, \bibinfo {author}
  {\bibfnamefont {J.~H.}\ \bibnamefont {Yoo}}, \emph {et~al.},\ }\bibfield
  {title} {\bibinfo {title} {Tensor{F}low {Q}uantum: A software framework for
  quantum machine learning},\ }\href@noop {} {\bibfield  {journal} {\bibinfo
  {journal} {arXiv:2003.02989}\ } (\bibinfo {year} {2020})}\BibitemShut
  {NoStop}%
\bibitem [{\citenamefont {Perelshtein}\ and\ \citenamefont
  {Pakhomchik}(2021)}]{QuEnc_patent}%
  \BibitemOpen
  \bibfield  {author} {\bibinfo {author} {\bibfnamefont {M.~R.}\ \bibnamefont
  {Perelshtein}}\ and\ \bibinfo {author} {\bibfnamefont {A.~I.}\ \bibnamefont
  {Pakhomchik}},\ }\href@noop {} {\bibinfo {title} {Hardware-efficient hybrid
  quantum algorithm for discrete optimization}},\ \bibinfo {howpublished}
  {Patent} (\bibinfo {year} {2021})\BibitemShut {NoStop}%
\bibitem [{\citenamefont {Dunjko}\ and\ \citenamefont
  {Briegel}(2018)}]{dunjko2018machine}%
  \BibitemOpen
  \bibfield  {author} {\bibinfo {author} {\bibfnamefont {V.}~\bibnamefont
  {Dunjko}}\ and\ \bibinfo {author} {\bibfnamefont {H.~J.}\ \bibnamefont
  {Briegel}},\ }\bibfield  {title} {\bibinfo {title} {Machine learning \&
  artificial intelligence in the quantum domain: a review of recent progress},\
  }\href@noop {} {\bibfield  {journal} {\bibinfo  {journal} {Rep. Prog. Phys.}\
  }\textbf {\bibinfo {volume} {81}},\ \bibinfo {pages} {074001} (\bibinfo
  {year} {2018})}\BibitemShut {NoStop}%
\bibitem [{\citenamefont {Carleo}\ \emph {et~al.}(2019)\citenamefont {Carleo},
  \citenamefont {Cirac}, \citenamefont {Cranmer}, \citenamefont {Daudet},
  \citenamefont {Schuld}, \citenamefont {Tishby}, \citenamefont
  {Vogt-Maranto},\ and\ \citenamefont {Zdeborov\'a}}]{carleo2019machine}%
  \BibitemOpen
  \bibfield  {author} {\bibinfo {author} {\bibfnamefont {G.}~\bibnamefont
  {Carleo}}, \bibinfo {author} {\bibfnamefont {I.}~\bibnamefont {Cirac}},
  \bibinfo {author} {\bibfnamefont {K.}~\bibnamefont {Cranmer}}, \bibinfo
  {author} {\bibfnamefont {L.}~\bibnamefont {Daudet}}, \bibinfo {author}
  {\bibfnamefont {M.}~\bibnamefont {Schuld}}, \bibinfo {author} {\bibfnamefont
  {N.}~\bibnamefont {Tishby}}, \bibinfo {author} {\bibfnamefont
  {L.}~\bibnamefont {Vogt-Maranto}},\ and\ \bibinfo {author} {\bibfnamefont
  {L.}~\bibnamefont {Zdeborov\'a}},\ }\bibfield  {title} {\bibinfo {title}
  {Machine learning and the physical sciences},\ }\href@noop {} {\bibfield
  {journal} {\bibinfo  {journal} {Rev. Mod. Phys.}\ }\textbf {\bibinfo {volume}
  {91}},\ \bibinfo {pages} {045002} (\bibinfo {year} {2019})}\BibitemShut
  {NoStop}%
\bibitem [{\citenamefont {Biamonte}\ \emph {et~al.}(2017)\citenamefont
  {Biamonte}, \citenamefont {Wittek}, \citenamefont {Pancotti}, \citenamefont
  {Rebentrost}, \citenamefont {Wiebe},\ and\ \citenamefont
  {Lloyd}}]{biamonte2017quantum}%
  \BibitemOpen
  \bibfield  {author} {\bibinfo {author} {\bibfnamefont {J.}~\bibnamefont
  {Biamonte}}, \bibinfo {author} {\bibfnamefont {P.}~\bibnamefont {Wittek}},
  \bibinfo {author} {\bibfnamefont {N.}~\bibnamefont {Pancotti}}, \bibinfo
  {author} {\bibfnamefont {P.}~\bibnamefont {Rebentrost}}, \bibinfo {author}
  {\bibfnamefont {N.}~\bibnamefont {Wiebe}},\ and\ \bibinfo {author}
  {\bibfnamefont {S.}~\bibnamefont {Lloyd}},\ }\bibfield  {title} {\bibinfo
  {title} {Quantum machine learning},\ }\href@noop {} {\bibfield  {journal}
  {\bibinfo  {journal} {Nature}\ }\textbf {\bibinfo {volume} {549}},\ \bibinfo
  {pages} {195} (\bibinfo {year} {2017})}\BibitemShut {NoStop}%
\bibitem [{\citenamefont {F\"osel}\ \emph {et~al.}(2018)\citenamefont
  {F\"osel}, \citenamefont {Tighineanu}, \citenamefont {Weiss},\ and\
  \citenamefont {Marquardt}}]{fosel2018reinforcement}%
  \BibitemOpen
  \bibfield  {author} {\bibinfo {author} {\bibfnamefont {T.}~\bibnamefont
  {F\"osel}}, \bibinfo {author} {\bibfnamefont {P.}~\bibnamefont {Tighineanu}},
  \bibinfo {author} {\bibfnamefont {T.}~\bibnamefont {Weiss}},\ and\ \bibinfo
  {author} {\bibfnamefont {F.}~\bibnamefont {Marquardt}},\ }\bibfield  {title}
  {\bibinfo {title} {Reinforcement learning with neural networks for quantum
  feedback},\ }\href@noop {} {\bibfield  {journal} {\bibinfo  {journal} {Phys.
  Rev. X}\ }\textbf {\bibinfo {volume} {8}},\ \bibinfo {pages} {031084}
  (\bibinfo {year} {2018})}\BibitemShut {NoStop}%
\bibitem [{\citenamefont {Bukov}\ \emph {et~al.}(2018)\citenamefont {Bukov},
  \citenamefont {Day}, \citenamefont {Sels}, \citenamefont {Weinberg},
  \citenamefont {Polkovnikov},\ and\ \citenamefont
  {Mehta}}]{bukov2018reinforcement}%
  \BibitemOpen
  \bibfield  {author} {\bibinfo {author} {\bibfnamefont {M.}~\bibnamefont
  {Bukov}}, \bibinfo {author} {\bibfnamefont {A.~G.~R.}\ \bibnamefont {Day}},
  \bibinfo {author} {\bibfnamefont {D.}~\bibnamefont {Sels}}, \bibinfo {author}
  {\bibfnamefont {P.}~\bibnamefont {Weinberg}}, \bibinfo {author}
  {\bibfnamefont {A.}~\bibnamefont {Polkovnikov}},\ and\ \bibinfo {author}
  {\bibfnamefont {P.}~\bibnamefont {Mehta}},\ }\bibfield  {title} {\bibinfo
  {title} {Reinforcement learning in different phases of quantum control},\
  }\href@noop {} {\bibfield  {journal} {\bibinfo  {journal} {Phys. Rev. X}\
  }\textbf {\bibinfo {volume} {8}},\ \bibinfo {pages} {031086} (\bibinfo {year}
  {2018})}\BibitemShut {NoStop}%
\bibitem [{\citenamefont {Xu}\ \emph {et~al.}(2019)\citenamefont {Xu},
  \citenamefont {Li}, \citenamefont {Liu}, \citenamefont {Wang}, \citenamefont
  {Yuan},\ and\ \citenamefont {Wang}}]{xu2019generalizable}%
  \BibitemOpen
  \bibfield  {author} {\bibinfo {author} {\bibfnamefont {H.}~\bibnamefont
  {Xu}}, \bibinfo {author} {\bibfnamefont {J.}~\bibnamefont {Li}}, \bibinfo
  {author} {\bibfnamefont {L.}~\bibnamefont {Liu}}, \bibinfo {author}
  {\bibfnamefont {Y.}~\bibnamefont {Wang}}, \bibinfo {author} {\bibfnamefont
  {H.}~\bibnamefont {Yuan}},\ and\ \bibinfo {author} {\bibfnamefont
  {X.}~\bibnamefont {Wang}},\ }\bibfield  {title} {\bibinfo {title}
  {Generalizable control for quantum parameter estimation through reinforcement
  learning},\ }\href@noop {} {\bibfield  {journal} {\bibinfo  {journal} {npj
  Quantum Inf.}\ }\textbf {\bibinfo {volume} {5}},\ \bibinfo {pages} {82}
  (\bibinfo {year} {2019})}\BibitemShut {NoStop}%
\bibitem [{\citenamefont {Poulsen~Nautrup}\ \emph {et~al.}(2019)\citenamefont
  {Poulsen~Nautrup}, \citenamefont {Delfosse}, \citenamefont {Dunjko},
  \citenamefont {Briegel},\ and\ \citenamefont
  {Friis}}]{nautrup2019optimizing}%
  \BibitemOpen
  \bibfield  {author} {\bibinfo {author} {\bibfnamefont {H.}~\bibnamefont
  {Poulsen~Nautrup}}, \bibinfo {author} {\bibfnamefont {N.}~\bibnamefont
  {Delfosse}}, \bibinfo {author} {\bibfnamefont {V.}~\bibnamefont {Dunjko}},
  \bibinfo {author} {\bibfnamefont {H.~J.}\ \bibnamefont {Briegel}},\ and\
  \bibinfo {author} {\bibfnamefont {N.}~\bibnamefont {Friis}},\ }\bibfield
  {title} {\bibinfo {title} {Optimizing quantum error correction codes with
  reinforcement learning},\ }\href@noop {} {\bibfield  {journal} {\bibinfo
  {journal} {{Quantum}}\ }\textbf {\bibinfo {volume} {3}},\ \bibinfo {pages}
  {215} (\bibinfo {year} {2019})}\BibitemShut {NoStop}%
\bibitem [{\citenamefont {Sweke}\ \emph {et~al.}(2020)\citenamefont {Sweke},
  \citenamefont {Kesselring}, \citenamefont {van Nieuwenburg},\ and\
  \citenamefont {Eisert}}]{sweke2020reinforcement}%
  \BibitemOpen
  \bibfield  {author} {\bibinfo {author} {\bibfnamefont {R.}~\bibnamefont
  {Sweke}}, \bibinfo {author} {\bibfnamefont {M.~S.}\ \bibnamefont
  {Kesselring}}, \bibinfo {author} {\bibfnamefont {E.~P.}\ \bibnamefont {van
  Nieuwenburg}},\ and\ \bibinfo {author} {\bibfnamefont {J.}~\bibnamefont
  {Eisert}},\ }\bibfield  {title} {\bibinfo {title} {Reinforcement learning
  decoders for fault-tolerant quantum computation},\ }\href@noop {} {\bibfield
  {journal} {\bibinfo  {journal} {Mach. Learn.: Sci. Technol.}\ }\textbf
  {\bibinfo {volume} {2}},\ \bibinfo {pages} {025005} (\bibinfo {year}
  {2020})}\BibitemShut {NoStop}%
\bibitem [{\citenamefont {Melnikov}\ \emph {et~al.}(2018)\citenamefont
  {Melnikov}, \citenamefont {Poulsen~Nautrup}, \citenamefont {Krenn},
  \citenamefont {Dunjko}, \citenamefont {Tiersch}, \citenamefont {Zeilinger},\
  and\ \citenamefont {Briegel}}]{melnikov2018active}%
  \BibitemOpen
  \bibfield  {author} {\bibinfo {author} {\bibfnamefont {A.~A.}\ \bibnamefont
  {Melnikov}}, \bibinfo {author} {\bibfnamefont {H.}~\bibnamefont
  {Poulsen~Nautrup}}, \bibinfo {author} {\bibfnamefont {M.}~\bibnamefont
  {Krenn}}, \bibinfo {author} {\bibfnamefont {V.}~\bibnamefont {Dunjko}},
  \bibinfo {author} {\bibfnamefont {M.}~\bibnamefont {Tiersch}}, \bibinfo
  {author} {\bibfnamefont {A.}~\bibnamefont {Zeilinger}},\ and\ \bibinfo
  {author} {\bibfnamefont {H.~J.}\ \bibnamefont {Briegel}},\ }\bibfield
  {title} {\bibinfo {title} {Active learning machine learns to create new
  quantum experiments},\ }\href@noop {} {\bibfield  {journal} {\bibinfo
  {journal} {Proc. Natl. Acad. Sci. U.S.A.}\ }\textbf {\bibinfo {volume}
  {115}},\ \bibinfo {pages} {1221} (\bibinfo {year} {2018})}\BibitemShut
  {NoStop}%
\bibitem [{\citenamefont {Walln\"ofer}\ \emph {et~al.}(2020)\citenamefont
  {Walln\"ofer}, \citenamefont {Melnikov}, \citenamefont {D\"ur},\ and\
  \citenamefont {Briegel}}]{wallnofer2020machine}%
  \BibitemOpen
  \bibfield  {author} {\bibinfo {author} {\bibfnamefont {J.}~\bibnamefont
  {Walln\"ofer}}, \bibinfo {author} {\bibfnamefont {A.~A.}\ \bibnamefont
  {Melnikov}}, \bibinfo {author} {\bibfnamefont {W.}~\bibnamefont {D\"ur}},\
  and\ \bibinfo {author} {\bibfnamefont {H.~J.}\ \bibnamefont {Briegel}},\
  }\bibfield  {title} {\bibinfo {title} {Machine learning for long-distance
  quantum communication},\ }\href@noop {} {\bibfield  {journal} {\bibinfo
  {journal} {PRX Quantum}\ }\textbf {\bibinfo {volume} {1}},\ \bibinfo {pages}
  {010301} (\bibinfo {year} {2020})}\BibitemShut {NoStop}%
\bibitem [{\citenamefont {Melnikov}\ \emph
  {et~al.}(2020{\natexlab{a}})\citenamefont {Melnikov}, \citenamefont
  {Sekatski},\ and\ \citenamefont {Sangouard}}]{melnikov2020setting}%
  \BibitemOpen
  \bibfield  {author} {\bibinfo {author} {\bibfnamefont {A.~A.}\ \bibnamefont
  {Melnikov}}, \bibinfo {author} {\bibfnamefont {P.}~\bibnamefont {Sekatski}},\
  and\ \bibinfo {author} {\bibfnamefont {N.}~\bibnamefont {Sangouard}},\
  }\bibfield  {title} {\bibinfo {title} {Setting up experimental {B}ell tests
  with reinforcement learning},\ }\href@noop {} {\bibfield  {journal} {\bibinfo
   {journal} {Phys. Rev. Lett.}\ }\textbf {\bibinfo {volume} {125}},\ \bibinfo
  {pages} {160401} (\bibinfo {year} {2020}{\natexlab{a}})}\BibitemShut
  {NoStop}%
\bibitem [{\citenamefont {Krenn}\ \emph {et~al.}(2016)\citenamefont {Krenn},
  \citenamefont {Malik}, \citenamefont {Fickler}, \citenamefont {Lapkiewicz},\
  and\ \citenamefont {Zeilinger}}]{krenn2016automated}%
  \BibitemOpen
  \bibfield  {author} {\bibinfo {author} {\bibfnamefont {M.}~\bibnamefont
  {Krenn}}, \bibinfo {author} {\bibfnamefont {M.}~\bibnamefont {Malik}},
  \bibinfo {author} {\bibfnamefont {R.}~\bibnamefont {Fickler}}, \bibinfo
  {author} {\bibfnamefont {R.}~\bibnamefont {Lapkiewicz}},\ and\ \bibinfo
  {author} {\bibfnamefont {A.}~\bibnamefont {Zeilinger}},\ }\bibfield  {title}
  {\bibinfo {title} {Automated search for new quantum experiments},\
  }\href@noop {} {\bibfield  {journal} {\bibinfo  {journal} {Phys. Rev. Lett.}\
  }\textbf {\bibinfo {volume} {116}},\ \bibinfo {pages} {090405} (\bibinfo
  {year} {2016})}\BibitemShut {NoStop}%
\bibitem [{\citenamefont {Krenn}\ \emph {et~al.}(2020)\citenamefont {Krenn},
  \citenamefont {Erhard},\ and\ \citenamefont {Zeilinger}}]{krenn2020computer}%
  \BibitemOpen
  \bibfield  {author} {\bibinfo {author} {\bibfnamefont {M.}~\bibnamefont
  {Krenn}}, \bibinfo {author} {\bibfnamefont {M.}~\bibnamefont {Erhard}},\ and\
  \bibinfo {author} {\bibfnamefont {A.}~\bibnamefont {Zeilinger}},\ }\bibfield
  {title} {\bibinfo {title} {Computer-inspired quantum experiments},\
  }\href@noop {} {\bibfield  {journal} {\bibinfo  {journal} {Nat. Rev. Phys.}\
  }\textbf {\bibinfo {volume} {2}},\ \bibinfo {pages} {649} (\bibinfo {year}
  {2020})}\BibitemShut {NoStop}%
\bibitem [{\citenamefont {Melnikov}\ \emph {et~al.}(2019)\citenamefont
  {Melnikov}, \citenamefont {Fedichkin},\ and\ \citenamefont
  {Alodjants}}]{melnikov2019predicting}%
  \BibitemOpen
  \bibfield  {author} {\bibinfo {author} {\bibfnamefont {A.~A.}\ \bibnamefont
  {Melnikov}}, \bibinfo {author} {\bibfnamefont {L.~E.}\ \bibnamefont
  {Fedichkin}},\ and\ \bibinfo {author} {\bibfnamefont {A.}~\bibnamefont
  {Alodjants}},\ }\bibfield  {title} {\bibinfo {title} {Predicting quantum
  advantage by quantum walk with convolutional neural networks},\ }\href@noop
  {} {\bibfield  {journal} {\bibinfo  {journal} {New J. Phys.}\ }\textbf
  {\bibinfo {volume} {21}},\ \bibinfo {pages} {125002} (\bibinfo {year}
  {2019})}\BibitemShut {NoStop}%
\bibitem [{\citenamefont {Melnikov}\ \emph
  {et~al.}(2020{\natexlab{b}})\citenamefont {Melnikov}, \citenamefont
  {Fedichkin}, \citenamefont {Lee},\ and\ \citenamefont
  {Alodjants}}]{melnikov2020machinetransfer}%
  \BibitemOpen
  \bibfield  {author} {\bibinfo {author} {\bibfnamefont {A.~A.}\ \bibnamefont
  {Melnikov}}, \bibinfo {author} {\bibfnamefont {L.~E.}\ \bibnamefont
  {Fedichkin}}, \bibinfo {author} {\bibfnamefont {R.-K.}\ \bibnamefont {Lee}},\
  and\ \bibinfo {author} {\bibfnamefont {A.}~\bibnamefont {Alodjants}},\
  }\bibfield  {title} {\bibinfo {title} {Machine learning transfer efficiencies
  for noisy quantum walks},\ }\href@noop {} {\bibfield  {journal} {\bibinfo
  {journal} {Adv. Quantum Technol.}\ }\textbf {\bibinfo {volume} {3}},\
  \bibinfo {pages} {1900115} (\bibinfo {year}
  {2020}{\natexlab{b}})}\BibitemShut {NoStop}%
\bibitem [{\citenamefont {Moussa}\ \emph {et~al.}(2020)\citenamefont {Moussa},
  \citenamefont {Calandra},\ and\ \citenamefont {Dunjko}}]{moussa2020quantum}%
  \BibitemOpen
  \bibfield  {author} {\bibinfo {author} {\bibfnamefont {C.}~\bibnamefont
  {Moussa}}, \bibinfo {author} {\bibfnamefont {H.}~\bibnamefont {Calandra}},\
  and\ \bibinfo {author} {\bibfnamefont {V.}~\bibnamefont {Dunjko}},\
  }\bibfield  {title} {\bibinfo {title} {To quantum or not to quantum: towards
  algorithm selection in near-term quantum optimization},\ }\href@noop {}
  {\bibfield  {journal} {\bibinfo  {journal} {Quantum Sci. Technol.}\ }\textbf
  {\bibinfo {volume} {5}},\ \bibinfo {pages} {044009} (\bibinfo {year}
  {2020})}\BibitemShut {NoStop}%
\bibitem [{\citenamefont {Yu}\ \emph {et~al.}(2019)\citenamefont {Yu},
  \citenamefont {Albarrán-Arriagada}, \citenamefont {Retamal}, \citenamefont
  {Wang}, \citenamefont {Liu}, \citenamefont {Ke}, \citenamefont {Meng},
  \citenamefont {Li}, \citenamefont {Tang}, \citenamefont {Solano},
  \citenamefont {Lamata}, \citenamefont {Li},\ and\ \citenamefont
  {Guo}}]{shang2019reconstruction}%
  \BibitemOpen
  \bibfield  {author} {\bibinfo {author} {\bibfnamefont {S.}~\bibnamefont
  {Yu}}, \bibinfo {author} {\bibfnamefont {F.}~\bibnamefont
  {Albarrán-Arriagada}}, \bibinfo {author} {\bibfnamefont {J.~C.}\
  \bibnamefont {Retamal}}, \bibinfo {author} {\bibfnamefont {Y.-T.}\
  \bibnamefont {Wang}}, \bibinfo {author} {\bibfnamefont {W.}~\bibnamefont
  {Liu}}, \bibinfo {author} {\bibfnamefont {Z.-J.}\ \bibnamefont {Ke}},
  \bibinfo {author} {\bibfnamefont {Y.}~\bibnamefont {Meng}}, \bibinfo {author}
  {\bibfnamefont {Z.-P.}\ \bibnamefont {Li}}, \bibinfo {author} {\bibfnamefont
  {J.-S.}\ \bibnamefont {Tang}}, \bibinfo {author} {\bibfnamefont
  {E.}~\bibnamefont {Solano}}, \bibinfo {author} {\bibfnamefont
  {L.}~\bibnamefont {Lamata}}, \bibinfo {author} {\bibfnamefont {C.-F.}\
  \bibnamefont {Li}},\ and\ \bibinfo {author} {\bibfnamefont {G.-C.}\
  \bibnamefont {Guo}},\ }\bibfield  {title} {\bibinfo {title} {Reconstruction
  of a photonic qubit state with reinforcement learning},\ }\href@noop {}
  {\bibfield  {journal} {\bibinfo  {journal} {Adv. Quantum Technol.}\ }\textbf
  {\bibinfo {volume} {2}},\ \bibinfo {pages} {1800074} (\bibinfo {year}
  {2019})}\BibitemShut {NoStop}%
\bibitem [{\citenamefont {Torlai}\ \emph {et~al.}(2019)\citenamefont {Torlai},
  \citenamefont {Timar}, \citenamefont {van Nieuwenburg}, \citenamefont
  {Levine}, \citenamefont {Omran} \emph {et~al.}}]{torlai2019integrating}%
  \BibitemOpen
  \bibfield  {author} {\bibinfo {author} {\bibfnamefont {G.}~\bibnamefont
  {Torlai}}, \bibinfo {author} {\bibfnamefont {B.}~\bibnamefont {Timar}},
  \bibinfo {author} {\bibfnamefont {E.~P.~L.}\ \bibnamefont {van Nieuwenburg}},
  \bibinfo {author} {\bibfnamefont {H.}~\bibnamefont {Levine}}, \bibinfo
  {author} {\bibfnamefont {A.}~\bibnamefont {Omran}}, \emph {et~al.},\
  }\bibfield  {title} {\bibinfo {title} {Integrating neural networks with a
  quantum simulator for state reconstruction},\ }\href@noop {} {\bibfield
  {journal} {\bibinfo  {journal} {Phys. Rev. Lett.}\ }\textbf {\bibinfo
  {volume} {123}},\ \bibinfo {pages} {230504} (\bibinfo {year}
  {2019})}\BibitemShut {NoStop}%
\bibitem [{\citenamefont {Palmieri}\ \emph {et~al.}(2020)\citenamefont
  {Palmieri}, \citenamefont {Kovlakov}, \citenamefont {Bianchi}, \citenamefont
  {Yudin}, \citenamefont {Straupe}, \citenamefont {Biamonte},\ and\
  \citenamefont {Kulik}}]{palmieri2020experimental}%
  \BibitemOpen
  \bibfield  {author} {\bibinfo {author} {\bibfnamefont {A.~M.}\ \bibnamefont
  {Palmieri}}, \bibinfo {author} {\bibfnamefont {E.}~\bibnamefont {Kovlakov}},
  \bibinfo {author} {\bibfnamefont {F.}~\bibnamefont {Bianchi}}, \bibinfo
  {author} {\bibfnamefont {D.}~\bibnamefont {Yudin}}, \bibinfo {author}
  {\bibfnamefont {S.}~\bibnamefont {Straupe}}, \bibinfo {author} {\bibfnamefont
  {J.~D.}\ \bibnamefont {Biamonte}},\ and\ \bibinfo {author} {\bibfnamefont
  {S.}~\bibnamefont {Kulik}},\ }\bibfield  {title} {\bibinfo {title}
  {Experimental neural network enhanced quantum tomography},\ }\href@noop {}
  {\bibfield  {journal} {\bibinfo  {journal} {npj Quantum Inf.}\ }\textbf
  {\bibinfo {volume} {6}} (\bibinfo {year} {2020})}\BibitemShut {NoStop}%
\bibitem [{\citenamefont {Ding}\ \emph {et~al.}(2020)\citenamefont {Ding},
  \citenamefont {Mart\'{\i}n-Guerrero}, \citenamefont {Sanz}, \citenamefont
  {Magdalena-Benedicto}, \citenamefont {Chen},\ and\ \citenamefont
  {Solano}}]{ding2020retrieving}%
  \BibitemOpen
  \bibfield  {author} {\bibinfo {author} {\bibfnamefont {Y.}~\bibnamefont
  {Ding}}, \bibinfo {author} {\bibfnamefont {J.~D.}\ \bibnamefont
  {Mart\'{\i}n-Guerrero}}, \bibinfo {author} {\bibfnamefont {M.}~\bibnamefont
  {Sanz}}, \bibinfo {author} {\bibfnamefont {R.}~\bibnamefont
  {Magdalena-Benedicto}}, \bibinfo {author} {\bibfnamefont {X.}~\bibnamefont
  {Chen}},\ and\ \bibinfo {author} {\bibfnamefont {E.}~\bibnamefont {Solano}},\
  }\bibfield  {title} {\bibinfo {title} {Retrieving quantum information with
  active learning},\ }\href@noop {} {\bibfield  {journal} {\bibinfo  {journal}
  {Phys. Rev. Lett.}\ }\textbf {\bibinfo {volume} {124}},\ \bibinfo {pages}
  {140504} (\bibinfo {year} {2020})}\BibitemShut {NoStop}%
\bibitem [{\citenamefont {Carleo}\ and\ \citenamefont
  {Troyer}(2017)}]{carleo2017solving}%
  \BibitemOpen
  \bibfield  {author} {\bibinfo {author} {\bibfnamefont {G.}~\bibnamefont
  {Carleo}}\ and\ \bibinfo {author} {\bibfnamefont {M.}~\bibnamefont
  {Troyer}},\ }\bibfield  {title} {\bibinfo {title} {Solving the quantum
  many-body problem with artificial neural networks},\ }\href@noop {}
  {\bibfield  {journal} {\bibinfo  {journal} {Science}\ }\textbf {\bibinfo
  {volume} {355}},\ \bibinfo {pages} {602} (\bibinfo {year}
  {2017})}\BibitemShut {NoStop}%
\bibitem [{\citenamefont {Gao}\ and\ \citenamefont
  {Duan}(2017)}]{gao2017efficient}%
  \BibitemOpen
  \bibfield  {author} {\bibinfo {author} {\bibfnamefont {X.}~\bibnamefont
  {Gao}}\ and\ \bibinfo {author} {\bibfnamefont {L.-M.}\ \bibnamefont {Duan}},\
  }\bibfield  {title} {\bibinfo {title} {Efficient representation of quantum
  many-body states with deep neural networks},\ }\href@noop {} {\bibfield
  {journal} {\bibinfo  {journal} {Nat. Commun.}\ }\textbf {\bibinfo {volume}
  {8}},\ \bibinfo {pages} {662} (\bibinfo {year} {2017})}\BibitemShut {NoStop}%
\bibitem [{\citenamefont {Le}\ \emph {et~al.}(2012)\citenamefont {Le},
  \citenamefont {Ranzato}, \citenamefont {Monga}, \citenamefont {Devin},
  \citenamefont {Chen}, \citenamefont {Corrado}, \citenamefont {Dean},\ and\
  \citenamefont {Ng}}]{Le2012}%
  \BibitemOpen
  \bibfield  {author} {\bibinfo {author} {\bibfnamefont {Q.~V.}\ \bibnamefont
  {Le}}, \bibinfo {author} {\bibfnamefont {M.}~\bibnamefont {Ranzato}},
  \bibinfo {author} {\bibfnamefont {R.}~\bibnamefont {Monga}}, \bibinfo
  {author} {\bibfnamefont {M.}~\bibnamefont {Devin}}, \bibinfo {author}
  {\bibfnamefont {K.}~\bibnamefont {Chen}}, \bibinfo {author} {\bibfnamefont
  {G.~S.}\ \bibnamefont {Corrado}}, \bibinfo {author} {\bibfnamefont
  {J.}~\bibnamefont {Dean}},\ and\ \bibinfo {author} {\bibfnamefont {A.~Y.}\
  \bibnamefont {Ng}},\ }\bibfield  {title} {\bibinfo {title} {Building
  high-level features using large scale unsupervised learning},\ }in\
  \href@noop {} {\emph {\bibinfo {booktitle} {29th Int. Conf. Mach. Learn.}}}\
  (\bibinfo {year} {2012})\BibitemShut {NoStop}%
\bibitem [{\citenamefont {Neven}\ \emph {et~al.}(2012)\citenamefont {Neven},
  \citenamefont {Denchev}, \citenamefont {Rose},\ and\ \citenamefont
  {Macready}}]{Neven2012QBoostLS}%
  \BibitemOpen
  \bibfield  {author} {\bibinfo {author} {\bibfnamefont {H.}~\bibnamefont
  {Neven}}, \bibinfo {author} {\bibfnamefont {V.~S.}\ \bibnamefont {Denchev}},
  \bibinfo {author} {\bibfnamefont {G.}~\bibnamefont {Rose}},\ and\ \bibinfo
  {author} {\bibfnamefont {W.~G.}\ \bibnamefont {Macready}},\ }\bibfield
  {title} {\bibinfo {title} {{QB}oost: Large scale classifier training
  withadiabatic quantum optimization},\ }in\ \href@noop {} {\emph {\bibinfo
  {booktitle} {Proc. Asian Conf. Mach. Learn.}}},\ \bibinfo {series}
  {Proceedings of Machine Learning Research}, Vol.~\bibinfo {volume} {25},\
  \bibinfo {editor} {edited by\ \bibinfo {editor} {\bibfnamefont {S.~C.~H.}\
  \bibnamefont {Hoi}}\ and\ \bibinfo {editor} {\bibfnamefont {W.}~\bibnamefont
  {Buntine}}}\ (\bibinfo  {publisher} {PMLR},\ \bibinfo {year} {2012})\ pp.\
  \bibinfo {pages} {333--348}\BibitemShut {NoStop}%
\bibitem [{\citenamefont {Rebentrost}\ \emph {et~al.}(2014)\citenamefont
  {Rebentrost}, \citenamefont {Mohseni},\ and\ \citenamefont
  {Lloyd}}]{PhysRevLett.113.130503}%
  \BibitemOpen
  \bibfield  {author} {\bibinfo {author} {\bibfnamefont {P.}~\bibnamefont
  {Rebentrost}}, \bibinfo {author} {\bibfnamefont {M.}~\bibnamefont
  {Mohseni}},\ and\ \bibinfo {author} {\bibfnamefont {S.}~\bibnamefont
  {Lloyd}},\ }\bibfield  {title} {\bibinfo {title} {Quantum support vector
  machine for big data classification},\ }\href@noop {} {\bibfield  {journal}
  {\bibinfo  {journal} {Phys. Rev. Lett.}\ }\textbf {\bibinfo {volume} {113}},\
  \bibinfo {pages} {130503} (\bibinfo {year} {2014})}\BibitemShut {NoStop}%
\bibitem [{\citenamefont {Saggio}\ \emph {et~al.}(2021)\citenamefont {Saggio},
  \citenamefont {Asenbeck}, \citenamefont {Hamann}, \citenamefont
  {Str{\"o}mberg}, \citenamefont {Schiansky}, \citenamefont {Dunjko},
  \citenamefont {Friis}, \citenamefont {Harris}, \citenamefont {Hochberg},
  \citenamefont {Englund} \emph {et~al.}}]{saggio2021experimental}%
  \BibitemOpen
  \bibfield  {author} {\bibinfo {author} {\bibfnamefont {V.}~\bibnamefont
  {Saggio}}, \bibinfo {author} {\bibfnamefont {B.~E.}\ \bibnamefont
  {Asenbeck}}, \bibinfo {author} {\bibfnamefont {A.}~\bibnamefont {Hamann}},
  \bibinfo {author} {\bibfnamefont {T.}~\bibnamefont {Str{\"o}mberg}}, \bibinfo
  {author} {\bibfnamefont {P.}~\bibnamefont {Schiansky}}, \bibinfo {author}
  {\bibfnamefont {V.}~\bibnamefont {Dunjko}}, \bibinfo {author} {\bibfnamefont
  {N.}~\bibnamefont {Friis}}, \bibinfo {author} {\bibfnamefont {N.~C.}\
  \bibnamefont {Harris}}, \bibinfo {author} {\bibfnamefont {M.}~\bibnamefont
  {Hochberg}}, \bibinfo {author} {\bibfnamefont {D.}~\bibnamefont {Englund}},
  \emph {et~al.},\ }\bibfield  {title} {\bibinfo {title} {Experimental quantum
  speed-up in reinforcement learning agents},\ }\href@noop {} {\bibfield
  {journal} {\bibinfo  {journal} {Nature}\ }\textbf {\bibinfo {volume} {591}},\
  \bibinfo {pages} {229} (\bibinfo {year} {2021})}\BibitemShut {NoStop}%
\bibitem [{\citenamefont {Lund}\ \emph {et~al.}(2017)\citenamefont {Lund},
  \citenamefont {Bremner},\ and\ \citenamefont {Ralph}}]{lund2017quantum}%
  \BibitemOpen
  \bibfield  {author} {\bibinfo {author} {\bibfnamefont {A.~P.}\ \bibnamefont
  {Lund}}, \bibinfo {author} {\bibfnamefont {M.~J.}\ \bibnamefont {Bremner}},\
  and\ \bibinfo {author} {\bibfnamefont {T.~C.}\ \bibnamefont {Ralph}},\
  }\bibfield  {title} {\bibinfo {title} {Quantum sampling problems,
  {B}oson{S}ampling and quantum supremacy},\ }\href@noop {} {\bibfield
  {journal} {\bibinfo  {journal} {npj Quantum Inf.}\ }\textbf {\bibinfo
  {volume} {3}},\ \bibinfo {pages} {1} (\bibinfo {year} {2017})}\BibitemShut
  {NoStop}%
\bibitem [{\citenamefont {Farhi}\ and\ \citenamefont {Neven}(2018)}]{Farhi}%
  \BibitemOpen
  \bibfield  {author} {\bibinfo {author} {\bibfnamefont {E.}~\bibnamefont
  {Farhi}}\ and\ \bibinfo {author} {\bibfnamefont {H.}~\bibnamefont {Neven}},\
  }\bibfield  {title} {\bibinfo {title} {Classification with quantum neural
  networks on near term processors},\ }\href@noop {} {\bibfield  {journal}
  {\bibinfo  {journal} {arXiv:1802.06002}\ } (\bibinfo {year}
  {2018})}\BibitemShut {NoStop}%
\bibitem [{\citenamefont {Rebentrost}\ \emph {et~al.}(2018)\citenamefont
  {Rebentrost}, \citenamefont {Bromley}, \citenamefont {Weedbrook},\ and\
  \citenamefont {Lloyd}}]{PhysRevA.98.042308}%
  \BibitemOpen
  \bibfield  {author} {\bibinfo {author} {\bibfnamefont {P.}~\bibnamefont
  {Rebentrost}}, \bibinfo {author} {\bibfnamefont {T.~R.}\ \bibnamefont
  {Bromley}}, \bibinfo {author} {\bibfnamefont {C.}~\bibnamefont {Weedbrook}},\
  and\ \bibinfo {author} {\bibfnamefont {S.}~\bibnamefont {Lloyd}},\ }\bibfield
   {title} {\bibinfo {title} {Quantum {H}opfield neural network},\ }\href@noop
  {} {\bibfield  {journal} {\bibinfo  {journal} {Phys. Rev. A}\ }\textbf
  {\bibinfo {volume} {98}},\ \bibinfo {pages} {042308} (\bibinfo {year}
  {2018})}\BibitemShut {NoStop}%
\bibitem [{\citenamefont {McClean}\ \emph {et~al.}(2018)\citenamefont
  {McClean}, \citenamefont {Boixo}, \citenamefont {Smelyanskiy}, \citenamefont
  {Babbush},\ and\ \citenamefont {Neven}}]{mcclean2018barren}%
  \BibitemOpen
  \bibfield  {author} {\bibinfo {author} {\bibfnamefont {J.~R.}\ \bibnamefont
  {McClean}}, \bibinfo {author} {\bibfnamefont {S.}~\bibnamefont {Boixo}},
  \bibinfo {author} {\bibfnamefont {V.~N.}\ \bibnamefont {Smelyanskiy}},
  \bibinfo {author} {\bibfnamefont {R.}~\bibnamefont {Babbush}},\ and\ \bibinfo
  {author} {\bibfnamefont {H.}~\bibnamefont {Neven}},\ }\bibfield  {title}
  {\bibinfo {title} {Barren plateaus in quantum neural network training
  landscapes},\ }\href@noop {} {\bibfield  {journal} {\bibinfo  {journal} {Nat.
  Commun.}\ }\textbf {\bibinfo {volume} {9}},\ \bibinfo {pages} {1} (\bibinfo
  {year} {2018})}\BibitemShut {NoStop}%
\bibitem [{\citenamefont {Beer}\ \emph {et~al.}(2020)\citenamefont {Beer},
  \citenamefont {Bondarenko}, \citenamefont {Farrelly}, \citenamefont
  {Osborne}, \citenamefont {Salzmann}, \citenamefont {Scheiermann},\ and\
  \citenamefont {Wolf}}]{beer2020training}%
  \BibitemOpen
  \bibfield  {author} {\bibinfo {author} {\bibfnamefont {K.}~\bibnamefont
  {Beer}}, \bibinfo {author} {\bibfnamefont {D.}~\bibnamefont {Bondarenko}},
  \bibinfo {author} {\bibfnamefont {T.}~\bibnamefont {Farrelly}}, \bibinfo
  {author} {\bibfnamefont {T.~J.}\ \bibnamefont {Osborne}}, \bibinfo {author}
  {\bibfnamefont {R.}~\bibnamefont {Salzmann}}, \bibinfo {author}
  {\bibfnamefont {D.}~\bibnamefont {Scheiermann}},\ and\ \bibinfo {author}
  {\bibfnamefont {R.}~\bibnamefont {Wolf}},\ }\bibfield  {title} {\bibinfo
  {title} {Training deep quantum neural networks},\ }\href@noop {} {\bibfield
  {journal} {\bibinfo  {journal} {Nat. Commun.}\ }\textbf {\bibinfo {volume}
  {11}},\ \bibinfo {pages} {1} (\bibinfo {year} {2020})}\BibitemShut {NoStop}%
\bibitem [{\citenamefont {Havl{\'{\i}}{\v{c}}ek}\ \emph
  {et~al.}(2019)\citenamefont {Havl{\'{\i}}{\v{c}}ek}, \citenamefont
  {C{\'{o}}rcoles}, \citenamefont {Temme}, \citenamefont {Harrow},
  \citenamefont {Kandala}, \citenamefont {Chow},\ and\ \citenamefont
  {Gambetta}}]{Havlek2019}%
  \BibitemOpen
  \bibfield  {author} {\bibinfo {author} {\bibfnamefont {V.}~\bibnamefont
  {Havl{\'{\i}}{\v{c}}ek}}, \bibinfo {author} {\bibfnamefont {A.~D.}\
  \bibnamefont {C{\'{o}}rcoles}}, \bibinfo {author} {\bibfnamefont
  {K.}~\bibnamefont {Temme}}, \bibinfo {author} {\bibfnamefont {A.~W.}\
  \bibnamefont {Harrow}}, \bibinfo {author} {\bibfnamefont {A.}~\bibnamefont
  {Kandala}}, \bibinfo {author} {\bibfnamefont {J.~M.}\ \bibnamefont {Chow}},\
  and\ \bibinfo {author} {\bibfnamefont {J.~M.}\ \bibnamefont {Gambetta}},\
  }\bibfield  {title} {\bibinfo {title} {Supervised learning with
  quantum-enhanced feature spaces},\ }\href@noop {} {\bibfield  {journal}
  {\bibinfo  {journal} {Nature}\ }\textbf {\bibinfo {volume} {567}},\ \bibinfo
  {pages} {209} (\bibinfo {year} {2019})}\BibitemShut {NoStop}%
\bibitem [{Cir(0 21)}]{CirclesDataset}%
  \BibitemOpen
  \href@noop {} {\bibinfo {title} {Scikit, {C}ircles {D}ataset}},\ \bibinfo
  {howpublished}
  {\url{https://scikit-learn.org/stable/modules/generated/sklearn.datasets.make_circles.html}}
  (\bibinfo {year} {Accessed: 2021-10-21})\BibitemShut {NoStop}%
\bibitem [{Bos(0 21)}]{BostonDataset}%
  \BibitemOpen
  \href@noop {} {\bibinfo {title} {Scikit, {B}oston {H}ousing {D}ataset}},\
  \bibinfo {howpublished}
  {\url{https://scikit-learn.org/stable/modules/generated/sklearn.datasets.load_boston.html}}
  (\bibinfo {year} {Accessed: 2021-10-21})\BibitemShut {NoStop}%
\bibitem [{\citenamefont {Harrison~Jr}\ and\ \citenamefont
  {Rubinfeld}(1978)}]{harrison1978hedonic}%
  \BibitemOpen
  \bibfield  {author} {\bibinfo {author} {\bibfnamefont {D.}~\bibnamefont
  {Harrison~Jr}}\ and\ \bibinfo {author} {\bibfnamefont {D.~L.}\ \bibnamefont
  {Rubinfeld}},\ }\bibfield  {title} {\bibinfo {title} {Hedonic housing prices
  and the demand for clean air},\ }\href@noop {} {\bibfield  {journal}
  {\bibinfo  {journal} {J. Environ. Econ. Manag.}\ }\textbf {\bibinfo {volume}
  {5}},\ \bibinfo {pages} {81} (\bibinfo {year} {1978})}\BibitemShut {NoStop}%
\bibitem [{\citenamefont {White}(1992)}]{DMRG}%
  \BibitemOpen
  \bibfield  {author} {\bibinfo {author} {\bibfnamefont {S.~R.}\ \bibnamefont
  {White}},\ }\bibfield  {title} {\bibinfo {title} {Density matrix formulation
  for quantum renormalization groups},\ }\href@noop {} {\bibfield  {journal}
  {\bibinfo  {journal} {Phys. Rev. Lett.}\ }\textbf {\bibinfo {volume} {69}},\
  \bibinfo {pages} {2863} (\bibinfo {year} {1992})}\BibitemShut {NoStop}%
\bibitem [{\citenamefont {Zhou}\ \emph {et~al.}(2020)\citenamefont {Zhou},
  \citenamefont {Stoudenmire},\ and\ \citenamefont {Waintal}}]{Zhou2020}%
  \BibitemOpen
  \bibfield  {author} {\bibinfo {author} {\bibfnamefont {Y.}~\bibnamefont
  {Zhou}}, \bibinfo {author} {\bibfnamefont {E.~M.}\ \bibnamefont
  {Stoudenmire}},\ and\ \bibinfo {author} {\bibfnamefont {X.}~\bibnamefont
  {Waintal}},\ }\bibfield  {title} {\bibinfo {title} {What limits the
  simulation of quantum computers?},\ }\href@noop {} {\bibfield  {journal}
  {\bibinfo  {journal} {Phys. Rev. X}\ }\textbf {\bibinfo {volume} {10}}
  (\bibinfo {year} {2020})}\BibitemShut {NoStop}%
\bibitem [{\citenamefont {Batchelor}(2000)}]{Batchelor2000}%
  \BibitemOpen
  \bibfield  {author} {\bibinfo {author} {\bibfnamefont {G.~K.}\ \bibnamefont
  {Batchelor}},\ }\bibfield  {title} {\bibinfo {title} {An introduction to
  fluid dynamics},\ }in\ \href@noop {} {\emph {\bibinfo {booktitle} {Cambridge
  University Press, Cambridge, UK}}}\ (\bibinfo {year} {2000})\BibitemShut
  {NoStop}%
\bibitem [{\citenamefont {Griffiths}(1999)}]{Griffiths1999}%
  \BibitemOpen
  \bibfield  {author} {\bibinfo {author} {\bibfnamefont {D.~J.}\ \bibnamefont
  {Griffiths}},\ }\bibfield  {title} {\bibinfo {title} {Introduction to
  {E}lectrodynamics},\ }in\ \href@noop {} {\emph {\bibinfo {booktitle}
  {Prentice Hall, Upper Saddle River, NJ}}}\ (\bibinfo {year}
  {1999})\BibitemShut {NoStop}%
\bibitem [{\citenamefont {Meyn}\ and\ \citenamefont
  {Tweedie}(2009)}]{Meyn2009}%
  \BibitemOpen
  \bibfield  {author} {\bibinfo {author} {\bibfnamefont {S.~P.}\ \bibnamefont
  {Meyn}}\ and\ \bibinfo {author} {\bibfnamefont {R.~L.}\ \bibnamefont
  {Tweedie}},\ }\bibfield  {title} {\bibinfo {title} {Markov chains and
  stochastic stability},\ }in\ \href@noop {} {\emph {\bibinfo {booktitle}
  {Cambridge University Press}}}\ (\bibinfo {year} {2009})\BibitemShut
  {NoStop}%
\bibitem [{\citenamefont {Engel}\ and\ \citenamefont
  {Dreizler}(2011)}]{Engel2011}%
  \BibitemOpen
  \bibfield  {author} {\bibinfo {author} {\bibfnamefont {E.}~\bibnamefont
  {Engel}}\ and\ \bibinfo {author} {\bibfnamefont {R.~M.}\ \bibnamefont
  {Dreizler}},\ }\bibfield  {title} {\bibinfo {title} {Density functional
  theory: An advanced course},\ }in\ \href@noop {} {\emph {\bibinfo {booktitle}
  {Springer, New York}}}\ (\bibinfo {year} {2011})\BibitemShut {NoStop}%
\bibitem [{\citenamefont {Kazeev}\ and\ \citenamefont
  {Khoromskij}(2012)}]{Kazeev2012}%
  \BibitemOpen
  \bibfield  {author} {\bibinfo {author} {\bibfnamefont {V.}~\bibnamefont
  {Kazeev}}\ and\ \bibinfo {author} {\bibfnamefont {B.}~\bibnamefont
  {Khoromskij}},\ }\bibfield  {title} {\bibinfo {title} {Low-rank explicit
  {QTT} representation of the laplace operator and its inverse},\ }\href@noop
  {} {\bibfield  {journal} {\bibinfo  {journal} {SIAM J. Matrix Anal. Appl.}\
  }\textbf {\bibinfo {volume} {33}} (\bibinfo {year} {2012})}\BibitemShut
  {NoStop}%
\bibitem [{\citenamefont {Dolgov}\ and\ \citenamefont
  {Savostyanov}(2014)}]{AMEN}%
  \BibitemOpen
  \bibfield  {author} {\bibinfo {author} {\bibfnamefont {S.}~\bibnamefont
  {Dolgov}}\ and\ \bibinfo {author} {\bibfnamefont {D.}~\bibnamefont
  {Savostyanov}},\ }\bibfield  {title} {\bibinfo {title} {Alternating minimal
  energy methods for linear systems in higher dimensions},\ }\href@noop {}
  {\bibfield  {journal} {\bibinfo  {journal} {SIAM J. Sci. Comput.}\ }\textbf
  {\bibinfo {volume} {36}},\ \bibinfo {pages} {1} (\bibinfo {year}
  {2014})}\BibitemShut {NoStop}%
\bibitem [{\citenamefont {Wang}\ \emph {et~al.}(2020)\citenamefont {Wang},
  \citenamefont {Wang}, \citenamefont {Li}, \citenamefont {Fan}, \citenamefont
  {Wei},\ and\ \citenamefont {Gu}}]{Wang2020}%
  \BibitemOpen
  \bibfield  {author} {\bibinfo {author} {\bibfnamefont {S.}~\bibnamefont
  {Wang}}, \bibinfo {author} {\bibfnamefont {Z.}~\bibnamefont {Wang}}, \bibinfo
  {author} {\bibfnamefont {W.}~\bibnamefont {Li}}, \bibinfo {author}
  {\bibfnamefont {L.}~\bibnamefont {Fan}}, \bibinfo {author} {\bibfnamefont
  {Z.}~\bibnamefont {Wei}},\ and\ \bibinfo {author} {\bibfnamefont
  {Y.}~\bibnamefont {Gu}},\ }\bibfield  {title} {\bibinfo {title} {Quantum fast
  {P}oisson solver: the algorithm and complete and modular circuit design},\
  }\href@noop {} {\bibfield  {journal} {\bibinfo  {journal} {Quantum Inf.
  Process.}\ }\textbf {\bibinfo {volume} {19}} (\bibinfo {year}
  {2020})}\BibitemShut {NoStop}%
\end{thebibliography}%

\end{document}